\documentclass[%
preprint,
preprintnumbers,
 amsmath,amssymb,
 aps,
 prfluids,
floatfix,
]{revtex4-2}

\usepackage{booktabs}
\usepackage{graphicx}     
\usepackage{dcolumn}      
\usepackage{bm}           
\usepackage{xcolor}
\usepackage{subcaption} 
\usepackage{algorithmic}
\usepackage{tikz}
\usetikzlibrary{shapes}
\usepackage[normalem]{ulem}
\usepackage[ruled,vlined]{algorithm2e}
\usepackage{wrapfig}
\usepackage{colortbl}

\usepackage{soul}
\usepackage{xparse}
\usepackage{etoolbox}

\colorlet{zycolor}{teal!50}

\NewDocumentCommand{\zy}{O{comment} m g}{%
  \sethlcolor{zycolor}%
  \IfNoValueTF{#3}{%
    \ifstrequal{#1}{comment}{\hl{#2}}{%
    \ifstrequal{#1}{delete}{\hl{\sout{#2}}}{%
    \ifstrequal{#1}{insert}{\hl{#2}}{%
      \PackageError{zy}{Unknown edit type `#1'}{}%
    }}}%
  }{%
    \ifstrequal{#1}{replace}{%
      \hl{\sout{#2} #3}%
    }{%
      \PackageError{zy}{Edit type `#1' does not take two arguments}{}%
    }%
  }%
}

\newcommand{\qw}[1]{{\color{black}{#1}}}

\usepackage{hyperref}
\hypersetup{
    colorlinks=true,
    linkcolor=blue,
    citecolor=blue,
    urlcolor=blue
}
\preprint{APS/123-QED}
\begin{document}

\title{Preconditioned Adjoint Data Assimilation for Two-Dimensional Decaying and Forced Turbulence}

\author{Hongyi Ke}
\author{Zejian You}
\author{Qi Wang}
\email{qwang4@sdsu.edu}
\affiliation{Department of Aerospace Engineering, San Diego State University, San Diego, CA 92182, USA}

\date{\today}

\begin{abstract}
Adjoint-based data assimilation for turbulent Navier-Stokes flows is fundamentally limited by the behavior of the adjoint dynamics: in backward time, adjoint fields exhibit exponential growth and become increasingly dominated by small-scale structures, severely degrading reconstruction of the initial condition from sparse measurements. We demonstrate that the relative weighting of spectral components in the adjoint formulation can be systematically controlled by redefining the inner product under which the adjoint operator is defined.
The inverse problem is formulated as a constrained minimization in which a cost functional measures the mismatch between model predictions and observations, and the adjoint equations provide the gradient with respect to the initial velocity field. Redefining the forward–adjoint duality through a Fourier-space weighting kernel preconditions the optimization and is mathematically equivalent to changing the representation of the control variable or, alternatively, introducing a smoothing operation on the governing dynamics. Specific kernel choices correspond to fractional integration or diffusion operators applied to the initial condition.
\qw{The method is demonstrated in two-dimensional forced Kolmogorov flow and decaying homogeneous isotropic turbulence. In the forced flow, the best
algebraic preconditioner reduces the initial-condition velocity error by more than an order of magnitude relative to the standard adjoint formulation, and
the preconditioning likewise improves the reconstruction in decaying turbulence. This gain is consistent with the wide separation between energy-containing scales and high-wavenumber amplification of the adjoint gradient, and the same mechanism, with a narrower separation, accounts for the result in decaying turbulence.} A statistical analysis of an ensemble of adjoint fields from different turbulent realizations reveals scale-dependent backward growth rates, explaining the instability of the standard formulation and clarifying the mechanism by which the proposed preconditioning attenuates incoherent small-scale amplification.
\end{abstract}

\maketitle
\section{Introduction}
\label{sec:introduction}
Reconstructing turbulent flows from sparse and noisy observations is a central and intrinsically challenging inverse problem in fluid dynamics \citep{mo2025reconstructing,angriman2023assimilation,zhang2021assimilation,duraisamy2019turbulence,zaki2025arfm,he2024four}, with applications in atmosphere-ocean prediction \citep{daley1997atmospheric,hamill2006ensemble}, closed-loop flow control \citep{ahmed2025efficient,ozan2025data}, and experimental flow measurements such as particle image velocimetry (PIV) \citep{gutierrez2026mixed,zhou2023flow,zhou2025joint}. The difficulty arises from the high dimensionality of the flow state, the nonlinear and chaotic dynamics governed by the Navier-Stokes equations, and the limited and indirect nature of practical measurements. These features render brute-force optimization or sampling approaches computationally intractable. Adjoint methods have therefore become the standard tool for PDE-constrained data assimilation, as they enable efficient evaluation of gradients of the cost functional with respect to high-dimensional control variables at a computational cost comparable to that of a forward simulation \citep{Dimet1986_4dvar,mahar1997optimal,courtier1997use}. Beyond gradient evaluation, adjoint-based sensitivity analysis continues to play an important role in data-driven flow reconstruction and in providing physical insight into flow receptivity and observability \citep{you2026localization,wang2025domain,wang_wang_zaki_2022}. Several fundamental questions remain unresolved, including the appropriate choice of inner product for defining adjoint duality for different flow setups \citep{scott2002optimal,protas2004computational,wang2025domain}, the integration of physical constraints with data-driven information, and the regularization of adjoint sensitivities in turbulent regimes \citep{garai2021stabilization}. Addressing these issues is essential before adjoint-based methods can be robustly deployed in practical high-Reynolds-number applications.

In the classical four-dimensional variational data assimilation(4D-Var) setting, the simulation tries to minimize the mismatch between the predicted measurements and the true observations. The gradient of the mismatch with respect to the initial condition or other control variables is obtained by solving the adjoint of the forward dynamics \citep{wang2019discrete, gutierrez2026mixed, gutierrez2026reconstructing}. The computational cost is approximately that of one forward and one adjoint solve, regardless of the number of control variables being optimized, in contrast to finite-difference sensitivities, whose cost scales linearly with the number of control variables \citep{Dimet1986_4dvar,mahar1997optimal,courtier1997use}.
For large-scale simulations, accurate gradient computation requires an adjoint consistent with the numerical discretization, including boundary conditions and mesh metrics, which has been derived and validated for many data-assimilation applications \citep{wang2019spatial, wang_wang_zaki_2022}. Practical implementations rely on checkpointing and incremental/outer-loop strategies to manage storage burden \citep{fisher2011long,courtier1997use}.
Besides evaluating the gradient, the sensitivity field calculated from adjoint Green's functions and its ensemble average \citep{wang2026mitigating} can be utilized to detect dangerous events with uncertainty quantification \citep{gutierrez2026reconstructing, you2026localization, you2026scalar, du2026inverse}, or optimize sensor placements \citep{mons_chassaing_sagaut_2017}.

When applied to turbulence, gradient-based approaches built upon the governing equations encounter fundamental difficulties. Turbulent flows are intrinsically multiscale and chaotic: nonlinear energy transfer populates a broad range of wavenumbers, and nearby initial conditions diverge exponentially in time. The adjoint system inherits this sensitivity but evolves backward in time. Consequently, perturbations aligned with the leading Lyapunov directions amplify exponentially and progressively concentrate at small scales \citep{zaki2025arfm,zaki2021prfaps}. Direct examinations of adjoint dynamics in wall-bounded and isotropic turbulence show rapid amplification of adjoint energy and steep spectral tilting toward high wavenumbers, yielding ill-conditioned gradients over long assimilation windows \citep{garai2021stabilization,nikitin2018characteristics}. This behavior is consistent with the Lyapunov spectrum of turbulent flows and explains why naive extensions of 4D-Var to long windows can diverge even when the forward solution is well resolved \citep{hassanaly2019lyapunov}. In practice, the resulting gradients overfit incoherent small-scale content, bias reconstructions, and mask the recovery of large-scale structures. Moreover, fundamental observability limits, e.g., from wall-based measurements, underline the difficulty of inferring full-field turbulence from limited data \citep{wang2021understanding,wang_wang_zaki_2022}. Together, these findings motivate scale-aware formulations that temper adjoint growth while preserving the physically relevant components of the sensitivity.

A variety of remedies have been proposed to temper adjoint growth while retaining reconstruction fidelity. Classical approaches impose \emph{a priori} smoothness or scale penalties on the control, e.g., Tikhonov/Sobolev regularization \citep{alberti2021learning,korn2019regularity}, and apply spectral filtering to suppress high-wavenumber noise. These approaches stabilize the optimization but may attenuate physically meaningful small scales or introduce bias. 
A related line of research has recognized that regularization can be achieved not only by augmenting the cost functional, but also by modifying the inner products (or ``brackets") used in the derivation of adjoint-based algorithms. In particular, \citet{scott2002optimal} pointed out that altering certain brackets appearing in the adjoint formulation may introduce a regularizing effect, leading to smoother and more numerically tractable gradients. Similar ideas were discussed in \citet{heinkenschloss1999interface} and later \citet{protas2004computational}.
Bayesian and ensemble-based schemes (e.g., hybrid/Ensemble Variational(EnVar) or Kriging-enhanced formulations) incorporate prior uncertainty and reduce sensitivity to local gradient pathologies at increased computational cost \citep{chandramoorthy2019feasibility,mons2019kriging,fisher2011long}.
It is also possible to change the solution space to effectively constrain the search within meaningful function space, constructed either by known prior distributions or through data-based methods, such as an auto-encoder \citep{wang2016auto, bauweraerts2021reconstruction}. \qw{Recently, \citet{cleary2025latent} and \citet{ozalp2026realtime} combine a learned latent representation with continuous and sequential data assimilation, respectively, for state estimation from sparse measurements. Their approaches restrict the state to a learned low-dimensional manifold, whereas ours retains the full state and the exact adjoint and changes the metric in which the gradient is expressed.}
Related stabilization ideas include synchronization/nudging and observability-aware formulations that respect fundamental limits of what can be inferred from restricted measurements \citep{wang_wang_zaki_2022,suzuki2017estimation}. \qw{For two-dimensional Kolmogorov flow, \citet{inubushi2026synchronisation} show that the essential observation resolution lies near the forcing scale, highlighting a strong separation between the observable and dissipative scales.} The open challenge is to design scale-aware procedures that simultaneously improve adjoint stability, maintain physical fidelity across wavenumbers, and remain computationally tractable for high-Reynolds-number flows.

Recent advances in scientific machine learning provide alternative strategies for such inverse problems in fluid mechanics. Physics-informed neural networks (PINNs) \citep{von2022mean,ehlers2025physics}, for example, incorporate governing equations as soft constraints in neural-network training and have been applied to data assimilation and flow reconstruction without explicitly forming adjoint equations. Similarly, operator-learning frameworks such as Deep Operator Network(DeepONet) \citep{hao2025assimilation} and Fourier neural operators \citep{moazzami2025meta} aim to approximate solution operators directly from data, enabling rapid inference once trained. More recently, diffusion and score-based generative models \citep{huang2024diffda,wang2025phyda,rozet2023score} have been adapted to physics-constrained inverse problems, where stochastic sampling in a learned latent space implicitly imposes a data-driven prior on the reconstructed fields. \qw{Relatedly, \citet{schiodt2026generative} apply stochastic interpolants patch-wise to super-resolve two-dimensional turbulence, recovering the energy spectrum and dissipation rate of the unresolved scales from low-resolution snapshots.} Closer to the present setting, machine learning has also been used to improve classical variational data assimilation rather than replace it. For example, \citet{frerix2021variational} use a learned model to initialize the non-convex 4D-Var optimization, while \citet{weyrauch2026state} combine machine learning with 4D-Var for state reconstruction in three-dimensional isotropic turbulence.

While these approaches have demonstrated promising reconstruction capabilities, they typically rely on learned representations whose spectral properties and stability characteristics are not explicitly linked to the dynamical structure of the Navier-Stokes equations. In particular, they do not directly address the backward Lyapunov growth \citep{eyink2004ruelle, wang_wang_zaki_2022, nikitin2018characteristics, wang2013forward} that underlies adjoint instability in turbulent flows, but rather mitigate ill-posedness through data-driven priors or stochastic regularization. Moreover, large-scale turbulent regimes remain challenging due to training cost, generalization limits, and the difficulty of embedding discretization-consistent numerical operators within deep architectures. These findings suggest that accurate reconstruction in turbulent flows requires both physical constraints and an informed treatment of how information is distributed across spatial scales.

The present work derives the preconditioned adjoint formulation, establishes the relationship between adjoint variables defined under different inner products, and recasts the optimization in terms of the transformed control variable. We also analyze the statistical and spectral characteristics of the adjoint sensitivity to identify its scale-dependent growth and to clarify the mechanism by which the kernel improves the inverse problem. \qw{The method is demonstrated for two-dimensional statistically stationary Kolmogorov flow and decaying homogeneous isotropic turbulence, with the standard and preconditioned formulations compared across several kernel choices.} The remainder of the paper is organized as follows. Section \ref{sec:adjoint} derives the adjoint equations within the data-assimilation framework. Section \ref{sec:inner_product} introduces generalized inner products and establishes the corresponding transformation between adjoint variables. In Section~\ref{sec:framework}, we show that this modification is equivalent to a change of control variable or, alternatively, to introducing a precursor smoothing operation, thereby formulating the preconditioned data-assimilation framework. \qw{Section \ref{sec:validation_control_variable} illustrates control-variable choices using Kolmogorov flow, and Section~\ref{sec:results_construction} introduces the spectral kernel families.}

Sections \ref{sec:results_power} and \ref{sec:results_exponential} investigate algebraic and exponential \qw{spectral preconditioning in forced Kolmogorov flow,} respectively, and compare their performance with the standard adjoint formulation. Section \qw{\ref{sec:results_hit}} applies the same \qw{preconditioner} families to two-dimensional \qw{decaying homogeneous isotropic turbulence and discusses the differences between the two flows.} \qw{Section~\ref{sec:sensitivity} examines the robustness of the results in both flows to the observation resolution and the length of the assimilation window.} Section \ref{sec:stats_adjoint} provides a statistical analysis of adjoint sensitivity for pointwise measurements, elucidating the scale-dependent growth mechanisms underlying the proposed stabilization for both flows. Concluding remarks and future directions are given in Section~\ref{sec:conclusion}.

\section{Mathematical Formulation}
\label{sec:formulation}

We first introduce the standard adjoint equations derived from the optimization using the forward two-dimensional Navier-Stokes equations as constraints, then derive the dependence of the adjoint gradient on the definition of the inner product, which is equivalent to preconditioning in the optimization procedure.

\subsection{Adjoint-based Data Assimilation}
\label{sec:adjoint}

Suppose the velocity field $\boldsymbol{u}(\boldsymbol{x},t)$ is a realization of two-dimensional turbulence, satisfying the two-dimensional incompressible Navier-Stokes equations in non-dimensional form,
\begin{equation}
\label{eq:NS}
\begin{aligned}
\frac{\partial \boldsymbol{u}}{\partial t}
 + \left(\boldsymbol{u}\cdot\nabla \right)\boldsymbol{u}
 &= - \nabla p + \nu \nabla^2 \boldsymbol{u}, \\
\nabla\cdot\boldsymbol{u} = 0 &, \quad\quad \boldsymbol{u}\big|_{t=0} = \boldsymbol{u}_0.
\end{aligned}
\end{equation}
Here, $\nu = \displaystyle\frac{1}{Re} = \frac{\displaystyle\nu^{\star}}{\displaystyle U^{\star} L^{\star}}$ denotes the non-dimensional kinematic viscosity. The superscript $\star$ denotes dimensional quantities, where $U^{\star}$, $L^{\star}$, and $\nu^{\star}$ are the dimensional characteristic velocity, length, and kinematic viscosity of the flow, respectively. The non-dimensional pressure is denoted as $p$. The initial condition
$\boldsymbol{u}(t=0) = \boldsymbol{u}_0$ is the goal of the data assimilation. 

Suppose the measurement kernel is denoted as a linear operator $\mathcal{M}$, the measurements can be obtained by applying the measurement kernel on the reference evolution of velocity, namely $\boldsymbol{m} = \mathcal{M}\left(\boldsymbol{u}_r\right)$, with $\boldsymbol{u}_r$ being the reference field solved with the true initial condition $\boldsymbol{u}_{0,r}$.
We construct the cost function as the norm of the mismatch between the modeled observation $\mathcal{M}\left(\boldsymbol{u}\right)$ and the true measurement $\boldsymbol{m}$,
\begin{equation}
    \mathcal{J} = \frac 12 \big\|\mathcal{M}\left(\boldsymbol{u}\right) - \boldsymbol{m}\big\|^2_2.
    \label{eq:J}
\end{equation}

The norm $\|\cdot\|_2$ in \eqref{eq:J} denotes the Euclidean norm for finite-dimensional vectors, consistent with $\mathcal{M}$ representing pointwise observations in space and time. Specifically, for observations at spatial locations $\{\boldsymbol{x}_i\}_{i=1}^{N_{\mathrm{obs}}}$ and times $\{t_j\}_{j=1}^{N_t}$,
\begin{equation}
\big\|\mathcal{M}(\boldsymbol{u})-\boldsymbol{m}\big\|_2^2
=
\sum_{j=1}^{N_t}\sum_{i=1}^{N_{\mathrm{obs}}}
\left|
\boldsymbol{u}(\boldsymbol{x}_i,t_j)-\boldsymbol{m}_{ij}
\right|^2,
\label{eq:J-norm}
\end{equation}
where \(|\cdot|\) denotes the Euclidean norm of the velocity vector at each observation point.


In some circumstances, we would like to utilize an 
augmented cost function by a Sobolev-type regularization on the initial condition to penalize high-wavenumber content and improve the conditioning of the inverse problem, namely
\begin{equation}
    \mathcal{J} \;=\; \frac{1}{2}\big\|\mathcal{M}\left(\boldsymbol{u}\right) - \boldsymbol{m}\big\|_2^2
      \;+\; \frac{\lambda}{2} \int_{\Omega} |\nabla \boldsymbol{u}_0|^2 \, d\Omega,
    \label{eq:J-reg}
\end{equation}
with $\lambda$ being a small positive parameter for the regularization.
Either way, the goal is to minimize the loss function $\mathcal{J}$ with equation \ref{eq:NS} as constraints.

We construct the Lagrangian using the Lagrange multiplier $\boldsymbol{u}^{\dagger}$ and $p^{\dagger}$, which are also the adjoint velocity and pressure,
\begin{equation}
    \mathcal{L} = \mathcal{J} - \int_0^T\left[ \frac{\partial \boldsymbol{u}}{\partial t}
 + \left(\boldsymbol{u}\cdot\nabla \right)\boldsymbol{u} + \nabla p - \nu \nabla^2 \boldsymbol{u}, \;\boldsymbol{u}^{\dagger} \right]dt - \int_0^T\left[ \nabla \cdot \boldsymbol{u}, p^{\dagger}\right]\;dt.
\end{equation}
Here the square brackets $\left[\cdot,\cdot\right]$ denote a spatial inner product defined as
\begin{equation}
\label{eqn:vanilla_inner_product}
    \left[\boldsymbol{a}, \boldsymbol{b}\right] =  \int_{\Omega} \boldsymbol{a}^{\top} \boldsymbol{b} \;d\Omega,
\end{equation}
for any vector- or scalar-valued fields $\boldsymbol{a}$ and $\boldsymbol{b}$.

When the initial condition $\boldsymbol{u}_0$ is perturbed, the linearized variations of the Lagrangian yield,
\begin{equation}
\label{eqn:variation}
    \delta \mathcal{L} = \delta \mathcal{J} - \int_0^T\bigg\{\left[ \frac{\partial \delta \boldsymbol{u}}{\partial t}
 + \left(\delta\boldsymbol{u}\cdot\nabla \right)\boldsymbol{u} + \left(\boldsymbol{u}\cdot\nabla \right)\delta\boldsymbol{u} + \nabla \delta p - \nu \nabla^2 \delta \boldsymbol{u}, \;\boldsymbol{u}^{\dagger} \right] + \left[ \nabla \cdot \delta\boldsymbol{u}, p^{\dagger}\right]\bigg\}dt.
\end{equation}
The perturbation of the cost function $\mathcal{J}$ can be expanded as,
\begin{equation}
    \label{eqn:perturbation_cost}
    \delta \mathcal{J} = \int_0^T\left[ \mathcal{M}^{\top}\left(\mathcal{M}(\boldsymbol{u}) - \boldsymbol{m}\right) - \delta(t)\;\lambda \nabla^2\boldsymbol{u}_0, \;\;\delta \boldsymbol{u} \right]\; dt.
\end{equation}
In the current study, $\mathcal{J}$ does not depend on pressure measurements. If pressure observations are included, an additional inner-product term involving $\delta p$ will be included in the variation.

Utilizing the spatial periodic boundary conditions, the inner product terms in equation \eqref{eqn:variation} can be rewritten using integration by parts as,
\begin{equation}
\label{eqn:integration_by_part}
\begin{aligned}
    &-\int_0^T \left\{\left[ \frac{\partial \delta \boldsymbol{u}}{\partial t}
 + \left(\delta\boldsymbol{u}\cdot\nabla \right)\boldsymbol{u} + \left(\boldsymbol{u}\cdot\nabla \right)\delta\boldsymbol{u} + \nabla \delta p - \nu\nabla^2 \delta \boldsymbol{u}, \;\boldsymbol{u}^{\dagger} \right] + \left[ \nabla \cdot \delta\boldsymbol{u}, p^{\dagger}\right]\right\}dt\\
 =& \left[ \boldsymbol{u}^{\dagger}_0,\; \delta\boldsymbol{u}_0\right] - \int_0^T \left\{\left[ \frac{\partial\boldsymbol{u}^{\dagger}}{\partial (-t)}
 +  \nabla \boldsymbol{u} \cdot \boldsymbol{u}^{\dagger} - \left( \boldsymbol{u}\cdot\nabla \right)\boldsymbol{u}^{\dagger} -  \nu\nabla^2 \boldsymbol{u}^{\dagger} - \nabla p^{\dagger}, \;\delta \boldsymbol{u} \right] - \left[ \nabla\cdot \boldsymbol{u}^{\dagger},\delta p \right] \right\}dt.
\end{aligned}
\end{equation}
The first term on the right arises from the boundary term in time integration. 
Combining the two expressions, we obtain,
\begin{equation}
    \delta \mathcal{L} =\int_0^T\left\{-\left[ \underbrace{\mathcal{N}^{\dagger}\boldsymbol{u}^{\dagger}}_{=\boldsymbol{0},\text{ Adjoint momentum eqns}}, \;\delta \boldsymbol{u} \right] + \left[ \underbrace{\nabla\cdot \boldsymbol{u}^{\dagger}}_{=0,\text{ Adjoint continuity}},\delta p \right]\right\}dt+ \left[ \boldsymbol{u}^{\dagger}_0,\; \delta\boldsymbol{u}_0\right].
\end{equation}

Here $\mathcal{N}^{\dagger}$ is the adjoint Navier-Stokes momentum operator. The adjoint equations read
\begin{equation}
\label{eq:AANS}
\begin{aligned}
\frac{\partial \boldsymbol{u}^\dagger}{\partial (-t)}
 + \nabla \boldsymbol{u}\cdot\boldsymbol{u}^\dagger
 - (\boldsymbol{u}\cdot\nabla)\boldsymbol{u}^\dagger 
 &= \nabla p^\dagger
    + \nu \nabla^2 \boldsymbol{u}^\dagger+ \mathcal{M}^{\top}\left(\mathcal{M}(\boldsymbol{u}) - \boldsymbol{m}\right) - \delta(t)\;\lambda \nabla^2\boldsymbol{u}_0, \\
\nabla\cdot\boldsymbol{u}^\dagger &= 0, \quad \boldsymbol{u}^{\dagger}\big\rvert_{t=T} = \boldsymbol{0}.
\end{aligned}
\end{equation}
The adjoint terminal condition $\boldsymbol{u}^{\dagger}\big\rvert_{t=T} = \boldsymbol{0}$ arises from integration by parts in time to eliminate the temporal boundary term.
The adjoint field may still be driven at the terminal time by the source term in the momentum equation, $\mathcal{M}^{\top}\left(\mathcal{M}(\boldsymbol{u}) - \boldsymbol{m}\right)$.

Setting the adjoint equations renders the relation between the variation of the Lagrangian $\mathcal{L}$ and the variation of the initial condition $\delta \boldsymbol{u}_0$ explicitly, and leads to the Fr\'echet derivative as the adjoint velocity at the initial time $t=0$,
\begin{equation}
    \delta \mathcal{L} = \left[ \boldsymbol{u}_0^{\dagger}, \; \delta \boldsymbol{u}_0\right], \quad \nabla_{\boldsymbol{u}_0}\mathcal{L} = \boldsymbol{u}^{\dagger}_0.
    \label{eqn:gradient_J}
\end{equation}
Here the gradient symbol $\nabla_{\boldsymbol{u}_0}$ denotes the Fr\'echet derivative using the inner product \eqref{eqn:vanilla_inner_product}.

\subsection{Choice of Inner Product and the Filtered Adjoint Variables}
\label{sec:inner_product}
It can be seen that the definition of the adjoint variable and the adjoint equations depend on the definition of the inner product. In general, we define a new inner product using the angle brackets,
\begin{equation}
\label{eqn:new_inner_product}
    \left\langle \boldsymbol{a}, \boldsymbol{b} \right\rangle
    = \int_{\Omega} \boldsymbol{a}^{\top} \mathcal{G}^{-1} \boldsymbol{b} \, d\Omega,
\end{equation}
for any vector- or scalar-valued fields $\boldsymbol{a}$ and $\boldsymbol{b}$.
Here $\mathcal{G} \succ 0$ is a symmetric, invertible convolution operator defined as
\begin{equation}
    \left(\mathcal{G}\boldsymbol{u}\right)(\boldsymbol{x}) = \int_{\Omega} G(\boldsymbol{x} - \boldsymbol{x}^{\prime})\boldsymbol{u}(\boldsymbol{x}^{\prime}) d\boldsymbol{x}^{\prime},
\end{equation}
with $\mathcal{G}^{-1}$ the inverse operation. The relation between this new inner product and the standard inner product $\left[\cdot,\cdot\right]$ can be summarized as,
\begin{equation}
\label{eqn:new_inner_product_relation}
    \left[ \boldsymbol{a}, \boldsymbol{b} \right]
    = \int_{\Omega} \boldsymbol{a}^{\top}  \boldsymbol{b} \, d\Omega = \int_{\Omega} \boldsymbol{a}^{\top}  \mathcal{G}^{-1} \mathcal{G}\,\boldsymbol{b} \, d\Omega = \left\langle \boldsymbol{a}, \mathcal{G}\boldsymbol{b}  \right\rangle = \left\langle \mathcal{G}\boldsymbol{a}, \boldsymbol{b}  \right\rangle,
\end{equation}
where symmetry of $\mathcal{G}$ has been used in the final equality.

If $G$ is a Dirac delta function, $\mathcal{G}$ and $\mathcal{G}^{-1}$ become identity maps, and the inner product $[\cdot,\cdot]$ becomes the same as $\left\langle \cdot ,\cdot \right\rangle$. 
In the current study, the convolution is assumed to be homogeneous in space, i.e., $G(\boldsymbol{x} - \boldsymbol{x}^{\prime})$ is only a function of the distance between $\boldsymbol{x}$ and $\boldsymbol{x}^{\prime}$.
In such cases, the convolution operator $\mathcal{G}$ and its inverse $\mathcal{G}^{-1}$ can be more conveniently defined in the wavenumber space as,
\begin{equation}
    \mathcal{G} = \mathcal{F}^{-1} \hat{G}(k)\mathcal{F},\qquad \mathcal{G}^{-1} = \mathcal{F}^{-1} \left(\hat{G}(k)\right)^{-1} \mathcal{F},
    \qquad
    k = |\boldsymbol{k}| = \sqrt{k_x^2 + k_y^2},
\end{equation}
where $\mathcal{F}$ and $\mathcal{F}^{-1}$ denote the Fourier transformation and its inverse. $\hat{G}$ is a real-valued, positive Fourier multiplier that depends only on $k$. Both $\mathcal{G}$ and $\mathcal{G}^{-1}$ act as modewise multipliers in Fourier space.
It is natural to choose $\mathcal{G}$ as a smoothing filter. In accordance with the standard notation in the turbulence literature, we denote the filtered field by $\tilde{\boldsymbol{u}} = \mathcal{G}\boldsymbol{u}$.

The resulting Fr\'echet derivative of the Lagrangian to the initial condition is, similar to \eqref{eqn:gradient_J}
\begin{equation}
    \delta \mathcal{L} = \left[ \boldsymbol{u}_0^{\dagger}, \; \delta \boldsymbol{u}_0\right] = \left\langle \mathcal{G}\boldsymbol{u}_0^{\dagger}, \; \delta \boldsymbol{u}_0\right\rangle= \left\langle \tilde{\boldsymbol{u}}_0^{\dagger}, \; \delta \boldsymbol{u}_0\right\rangle, \quad \nabla_{\boldsymbol{u}_0}^{\mathcal{G}}\mathcal{L} = \tilde{\boldsymbol{u}}^\dagger_0.
    \label{eqn:gradient_J2}
\end{equation}

Here $\tilde{\boldsymbol{u}}^{\dagger}=\mathcal{G}\boldsymbol{u}^{\dagger}$ denotes the filtered adjoint field, following the filtering convention $\tilde{\boldsymbol{u}}=\mathcal{G}\boldsymbol{u}$ introduced above. It is the adjoint gradient represented under the $\mathcal{G}$-weighted inner product. In particular,
$
\nabla_{\boldsymbol{u}_0}^{\mathcal{G}}\mathcal{L}
=
\tilde{\boldsymbol{u}}_0^{\dagger},
$
where $\tilde{\boldsymbol{u}}_0^{\dagger}$ denotes the filtered adjoint at the initial time.
The superscript $\mathcal{G}$ for the gradient symbol indicates that the gradient is defined with respect to the inner product induced by $\mathcal{G}$ in \eqref{eqn:new_inner_product}. It follows that the gradient is the \emph{filtered} adjoint velocity field at the initial time.
In principle, the filtered adjoint variable satisfies the corresponding filtered adjoint equations, i.e., a large-eddy-simulation (LES) form of the adjoint system. The derivation is written in Appendix \ref{app:LES_adjoint}, with its further implementation and analysis left for future investigation. The preconditioner modifies the inner product used in the optimization, resulting in the filtered gradient $\tilde{\boldsymbol{u}}_0^{\dagger}$ instead of the standard gradient $\boldsymbol{u}_0^{\dagger}$.

\subsection{Choice of Control Variable and Data Assimilation Framework}
\label{sec:framework}
Employing this gradient in \eqref{eqn:gradient_J2} requires that all inner products appearing in the optimization procedure be taken with respect to the modified inner product $\langle \cdot,\cdot \rangle$. In practice, this is not straightforward, as most standard optimization algorithms, e.g., \ Limited-memory Broyden-Fletcher-Goldfarb-Shanno (L-BFGS) method \citep{nocedal1980updating,liu1989limited}, implicitly rely on the Euclidean inner product when computing search directions and curvature information.
To avoid modifying the optimization algorithm itself, we instead introduce a transformed control variable $\boldsymbol{s}_0$. This transformation allows the optimization to be carried out in the standard Euclidean setting, while remaining equivalent to performing the optimization with respect to the weighted inner product in the original variable $\boldsymbol{u}_0$.

Notice that the inner product between two gradients defined by the $\nabla_{\boldsymbol{u}_0}^{\mathcal{G}}$ symbol is given by
\begin{equation}
    \left\langle \tilde{\boldsymbol{u}}^{\dagger(1)}_0, \tilde{\boldsymbol{u}}^{\dagger(2)}_0\right\rangle = \left\langle \mathcal{G}\boldsymbol{u}^{\dagger(1)}_0, \mathcal{G}\boldsymbol{u}^{\dagger(2)}_0\right\rangle= \left[ \mathcal{G}\boldsymbol{u}^{\dagger(1)}_0, \mathcal{G}^{-1}\mathcal{G}\boldsymbol{u}^{\dagger(2)}_0\right]
    =\left[ \underbrace{\mathcal{G}^{1/2}\boldsymbol{u}^{\dagger(1)}_0}_{\boldsymbol{s}_0^{\dagger(1)}},\; \underbrace{\mathcal{G}^{1/2}\boldsymbol{u}^{\dagger(2)}_0}_{\boldsymbol{s}_0^{\dagger(2)}}\right].
    \label{eqn:inner_product}
\end{equation}
Here $\mathcal{G}^{1/2} = \mathcal{F}^{-1} \left(\hat{G}(k)\right)^{1/2}\mathcal{F}$ is half a filter. This observation naturally motivates introducing a transformed control variable $\boldsymbol{s}_0$ such that 
\[
\nabla_{\boldsymbol{s}_0} \mathcal{L} = \boldsymbol{s}_0^{\dagger}=\mathcal{G}^{1/2}\boldsymbol{u}^{\dagger}_0.
\]
The $\boldsymbol{s}_0$ that satisfies this requirements is $\boldsymbol{s}_0 = \mathcal{G}^{-1/2}\boldsymbol{u}_0$, i.e., the ``half-sharpened initial condition". Since, by chain rule,
\begin{equation}
    \nabla_{\boldsymbol{s}_0} \mathcal{L} = \frac{\partial \boldsymbol{u}_0}{\partial \boldsymbol{s}_0} \nabla_{\boldsymbol{u}_0} \mathcal{L} = \mathcal{G}^{1/2} \boldsymbol{u}^\dagger_0 = \boldsymbol{s}^\dagger_0.
\label{eqn:gradientchange}
\end{equation}

\begin{figure}
    \centering
    \includegraphics[width=0.9\linewidth]{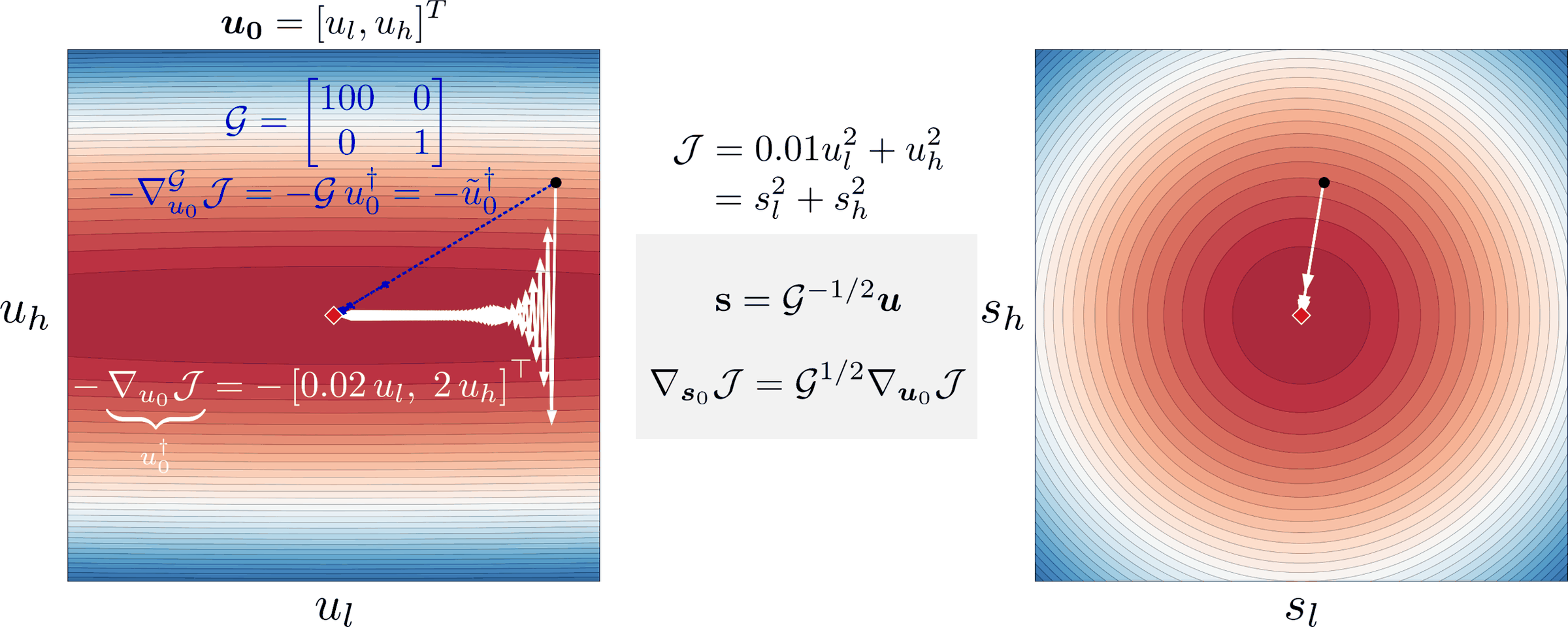}
    \caption{A toy problem demonstration of the change of inner product and the change of control variables on the impact of optimization. Subscripts $l$ and $h$ mark low- and high-sensitivity parts in the initial state $\boldsymbol{u}_0$.}
    \label{fig:toyproblem}
\end{figure}

This perspective establishes a precise mathematical connection between the choice of control variable, the definition of inner product, adjoint filtering, and preconditioning. 
To illustrate how the choice of inner product and control variable alters the effective geometry of the optimization, we consider a simple two-dimensional toy problem shown in figure \ref{fig:toyproblem}. We define the initial condition as the state vector $\boldsymbol{u}_0 = [u_l, u_h]^{\top}$, where the subscripts $l$ and $h$ denote low- and high-sensitivity directions, respectively. 
The cost function is defined as $\mathcal{J} = 0.01 u_l^2 + u_h^2$. In standard Euclidean space, the gradient (adjoint field) is given by $\nabla_{u_0} \mathcal{J}=  [0.02u_l, 2u_h]^{\top}$. The resulting steepest descent direction is heavily biased toward the high-sensitivity component, yielding a poorly conditioned, zig-zag path on the stretched level sets (left panel).
As demonstrated in Section~\ref{sec:inner_product}, the preconditioning can be viewed as a change of inner product. Choosing the filter $\mathcal{G} = \text{diag}(100, 1)$ compensates for the imbalance in sensitivity, producing a direct descent direction ($\nabla_{\boldsymbol{u}_0}^{\mathcal{G}} \mathcal{J}$, marked by the blue dashed arrow) in the original physical space. Equivalently, this can be framed as a change of control variable $\boldsymbol{s}_0 = \mathcal{G}^{-1/2}\boldsymbol{u}_0$. This transformation maps the distorted physical landscape to a latent space $\boldsymbol{s}_0$ (right panel) where the cost function becomes better conditioned in the new variable, i.e., $\mathcal{J} = s_l^2 + s_h^2$. Applying the chain rule, the gradient in the transformed variable is $\nabla_{\boldsymbol{s}_0}\mathcal{J} = \mathcal{G}^{1/2}\nabla_{\boldsymbol{u}_0}\mathcal{J} = [2s_l, 2s_h]^\top$, which yields a perfectly conditioned path to the minimizer and therefore requires fewer steps and less time to converge, while achieving higher accuracy.

\begin{figure}[h]
    \includegraphics[width=\textwidth]{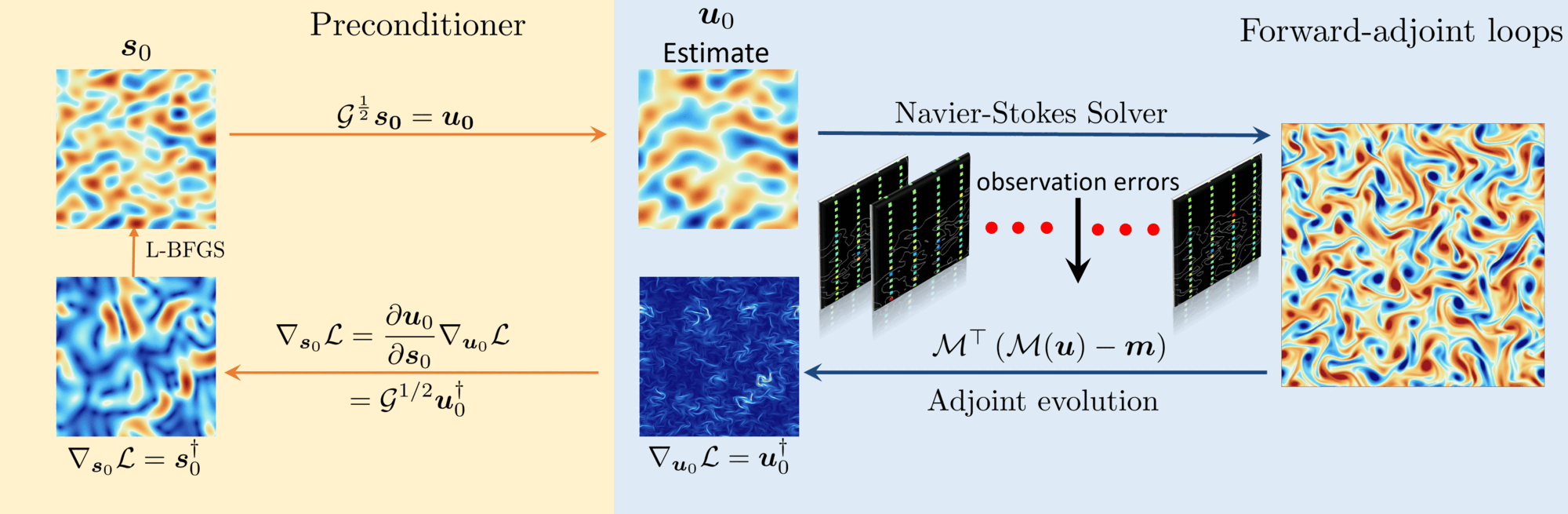}
    \caption{Data assimilation framework with preconditioner $\mathcal{G}$.}
    \label{fig:DA_framework}
\end{figure}

We illustrate this framework in figure \ref{fig:DA_framework}.
The observations consist of velocity measurements taken at spatial locations $\{\boldsymbol{x}_i^m\}_{i=1}^{N_{\mathrm{obs}}}$ and discrete times $\{t_j^m\}_{j=1}^{N_t}$. 
This corresponds to a measurement operator $\mathcal{M}$ composed of Dirac delta functions in space and time, mapping the model state to the observed velocity samples.
On the left side of figure \ref{fig:DA_framework}, we introduce the spectral preconditioner by redefining the control variable as
$\boldsymbol{s}_0 = \mathcal{G}^{-1/2}\boldsymbol{u}_0$.
If $\mathcal{G}$ is interpreted as a smoothing filter, then the adjoint gradient in the transformed space, 
$\boldsymbol{s}_0^{\dagger}$, corresponds to a filtered version of the physical adjoint $\boldsymbol{u}_0^{\dagger}$, while the mapping from $\boldsymbol{s}_0$ back to $\boldsymbol{u}_0$ introduces an additional filtering operation. 
Consequently, during each optimization step, the update direction applied to the physical initial condition effectively involves two operations of $\mathcal{G}^{1/2}$, resulting in a scale-selective damping of high-wavenumber components in the gradient.
The minimization of the cost functional $\mathcal{J}$ is carried out using the L-BFGS algorithm \citep{nocedal1980updating,liu1989limited}. It approximates the inverse Hessian from a limited history of gradients. Consequently, the optimization metric influences the approximate curvature seen by the algorithm. This is precisely where spectral preconditioning is effective: unlike an exact Newton method, whose iterates are independent of the choice of inner product, L-BFGS relies on an approximate Hessian, making its convergence sensitive to the metric used to define the gradient. 
For a linear dynamical system, with access to the exact Hessian of the quadratic cost function, such a preconditioner would not change the optimization direction. Its practical benefit here arises because turbulent data assimilation is nonlinear and the L-BFGS Hessian approximation is metric-dependent.
At iteration $k$, the control variable is updated according to
\[
\boldsymbol{s}_0^{(k+1)} 
\leftarrow 
\text{L-BFGS}\!\left(
\boldsymbol{s}_0^{(k)}, 
\boldsymbol{s}_0^{\dagger (k)}
\right),\quad k=1,\ldots N.
\]
after which the physical initial condition $\boldsymbol{u}_0$ is recovered through the $\mathcal{G}^{1/2}$ transformation.

This formulation is closely related to latent-space data assimilation \citep{cleary2025latent} and precursor-simulation strategies, in which the optimization is performed in a transformed representation of the state variables. 
In contrast to approaches that rely on extended backward integration windows to enhance reconstruction fidelity, the present method achieves scale-aware regularization directly through an explicit metric transformation in control space.

\section{Results}\label{sec:results}


\qw{We evaluate the proposed algorithm in two-dimensional forced Kolmogorov flow and decaying homogeneous isotropic turbulence (HIT).} \qw{Two-dimensional forced Kolmogorov flow is presented to illustrate control-variable choices and different choices of preconditioners.} \qw{We further examine the use of these preconditioners in a decaying HIT and compare their effectiveness.}
Numerical solutions are obtained via a fractional step method \citep{kim1985application}: an intermediate velocity field is computed by ignoring the continuity equation and pressure gradient in each time step, followed by solving the pressure Poisson equation and performing a projection of velocities onto a divergence-free space using the pressure gradient as the minimum correction. Temporal integration employs a second-order Adams-Bashforth scheme for the explicit advection and Crank-Nicolson for diffusion. Diffusion inversion and solving the pressure Poisson equation utilize modified wavenumbers in Fourier space, mirroring a central difference scheme.

\begin{table}[t]
\centering
\begin{tabular}{ccccccccccccc}
\toprule
 &
\multicolumn{2}{c}{Domain size} &
\multicolumn{2}{c}{Grid points} &
\multicolumn{2}{c}{Temporal resolution} &
\multicolumn{2}{c}{Observations} &
\multicolumn{2}{c}{Parameters} &
\multicolumn{2}{c}{Forcing} \\
\cmidrule(lr){2-3}\cmidrule(lr){4-5}\cmidrule(lr){6-7}
\cmidrule(lr){8-9}\cmidrule(lr){10-11}\cmidrule(lr){12-13}
 &
$L_x$ & $L_y$ &
$N_x$ & $N_y$ &
$\Delta t$ & $T$ &
$\Delta I$ & $\Delta t_m$ &
$Re$ & $N$ &
$n_f$ & $A$ \\
\midrule
HIT          & $2\pi$ & $2\pi$ & 512 & 512 & 0.002 & 2.0 & 8 & 0.10 & 1000 & 100 & --- & --- \\
Kolmogorov   & $2\pi$ & $2\pi$ & 512 & 512 & 0.001 & 2.0 & 8 & 0.05 &  400 & 100 & 4   & 1   \\
\bottomrule
\end{tabular}
\caption{Simulation and data-assimilation parameters for the 2D decaying HIT and Kolmogorov flow.}
\label{tab:sim_params}
\end{table}
\begin{figure}[ht]
    \centering
    \includegraphics[width=0.8\columnwidth]{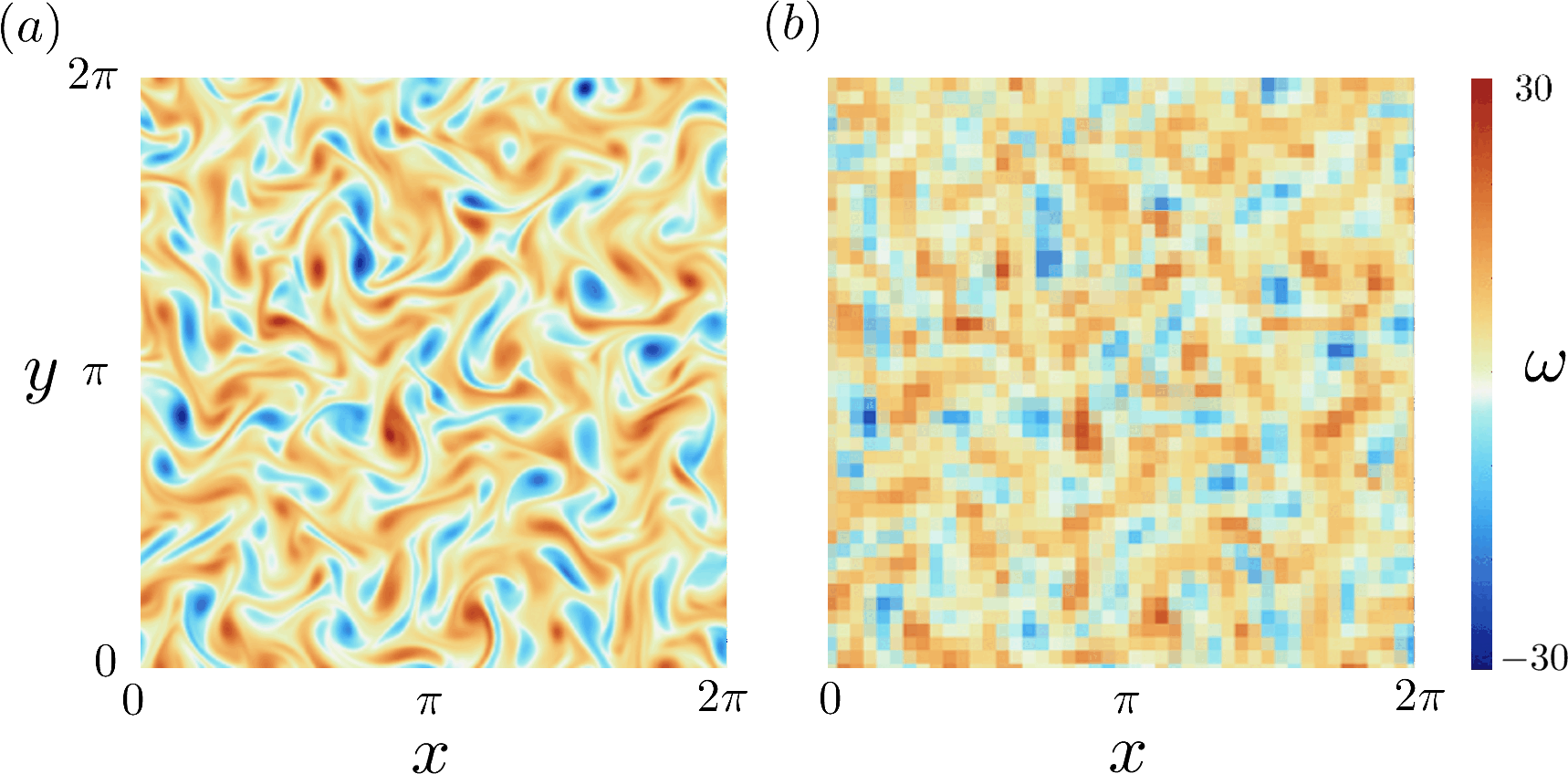}
    \caption{Illustration of the underlying grid resolution in a two-dimensional periodic domain of size $2\pi$. (a): full initial field on the $512\times512$ grid. (b): the observation obtained by downsampling with a grid gap of $\Delta I = 8$.}
    \label{fig:hyperparam_all}
\end{figure}
The adjoint solver is implemented through a summation-by-parts procedure, which ensures the gradient computations \eqref{eqn:gradient_J} and \eqref{eqn:gradient_J2} are satisfied to machine precision.
The simulation is performed on a double periodic domain $\Omega = [0, 2\pi)^2$, \qw{and the} relevant parameters are summarized in Table \ref{tab:sim_params}.
The one-dimensional (isotropic) energy spectrum $E(k)$ is defined from its two-dimensional Fourier transform,
\begin{equation}
E(k)
=
\sum_{k \le |\boldsymbol{k}| < k+\Delta k}
\underbrace{\frac 12 |\hat{\boldsymbol{u}}(\boldsymbol{k})|^2}_{\mathcal{E}(\boldsymbol{k})},
\end{equation}
where $k = |\boldsymbol{k}|$ and $\Delta k$ denotes the shell width in wavenumber space.
The initial condition for the two-dimensional decaying turbulence is generated from a prescribed isotropic energy spectrum of the form $E_{0,r}(k) = C\, k^{4}\exp\!\left[-\left(\frac{k}{k_0}\right)^2\right]$,
where $k_0$ denotes the characteristic energy-containing wavenumber, and $C$ is a normalization constant chosen to achieve the desired initial kinetic energy. Random phases are assigned in Fourier space, and the velocity field is projected onto a divergence-free space to ensure incompressibility before transforming back to physical space. The candidate field is advanced under the Navier--Stokes equations \eqref{eq:NS} for 0.2 time units to allow nonlinear interactions to develop a meaningful turbulent initial field, used as the truth initial condition $\boldsymbol{u}_{0,r}$ for the assimilation window of length $T=2$. 
A representative initial condition is shown in terms of vorticity contours in figure~\ref{fig:hyperparam_all}, which also illustrates the effect of the chosen observation gap $\Delta I=8$.

To characterize the difficulty of the reconstruction problem, Table~\ref{tab:physical_scales}
reports representative turbulence statistics of the two truth initial conditions,
including the \qw{Taylor microscale $\lambda_x$, Taylor Reynolds number
$Re_{\lambda,x}=u'_{\rm rms}\lambda_x/\nu$, and the Kraichnan dissipation
wavenumber
$k_\eta=(\chi/\nu^3)^{1/6}$ appropriate for two-dimensional turbulence.}
For HIT, $u'=u-\langle u\rangle_{xy}$. For Kolmogorov flow, the residual
velocity is obtained by removing the uniform mean and the four box-scale
modes $k_x^2+k_y^2=1$ from both velocity components.
The present decaying HIT case is comparable to the decaying
two-dimensional turbulence studied by \citet{jimenez2021collective}, which
also uses a random-phase initial field advanced through a transient before
analysis. Using the same box-scale Reynolds number definition,
$Re_L=q_{\rm rms}L/\nu$, where $q_{\rm rms}$ is the full-field two-component
r.m.s. speed, the truth field at the beginning of our assimilation
window gives $Re_L\approx5763$, within the range $Re_L=4400$--$11000$
reported in that study. This comparison provides a reference point for the
Reynolds-number regime of the present decaying-HIT experiment.


\qw{For both flows, $k_\eta$ lies well below the maximum resolved wavenumber $k_{\max}=256$, confirming that the $512^2$ Direct Numerical Simulation(DNS) adequately resolves the dissipation range.
For the Kolmogorov residual field, the 40 snapshots at $t=0.05,0.10,\ldots,2.00$
give $Re_{\lambda,x}=195\pm22$ (time mean and standard deviation).
These directional diagnostics characterize the sampled states.}

\begin{table}[t]
\centering
\qw{\begin{tabular}{lccccccc}
\toprule
 & $u'_{\rm rms}$ & $\lambda_x$ & $Re_{\lambda,x}$ & $\Omega$ & $\varepsilon_{\rm diss}$ & $\chi$ & $k_\eta$ 
 \\
\midrule
HIT ($Re=1000$)        & 0.647 & 0.128 & 83  & 99.6 & 0.199 & 108  & 69 
\\
Kolmogorov ($Re=400$)  & 0.801 & 0.612 & 196 & 8.37 & 0.042 & 1.13 & 20 
\\
\bottomrule
\end{tabular}}
\caption{Statistics of the truth initial conditions. The first three columns use the fluctuation/residual defined in the text: $u'_{\rm rms}=\langle u'^2\rangle_{xy}^{1/2}$, $\lambda_x^2=\langle u'^2\rangle_{xy}/\langle(\partial_xu')^2\rangle_{xy}$, and $Re_{\lambda,x}=u'_{\rm rms}\lambda_x/\nu$, with spatial averages over the domain. The remaining columns use the full field: $\Omega=\tfrac12\langle\omega^2\rangle$: enstrophy; $\varepsilon_{\rm diss}=\nu\langle|\nabla\boldsymbol{u}|^2\rangle$: energy dissipation rate; $\chi=\nu\langle|\nabla\omega|^2\rangle$: enstrophy dissipation rate; $k_\eta=(\chi/\nu^3)^{1/6}$: Kraichnan dissipation wavenumber. }
\label{tab:physical_scales}
\end{table}

\qw{For the forced Kolmogorov case, a deterministic body force $\boldsymbol{f}=(A\sin(n_f y),0)$ is added to the momentum equations,}
\begin{equation}
\label{eq:NS_kolm}
\begin{aligned}
\frac{\partial \boldsymbol{u}}{\partial t}
 + \left(\boldsymbol{u}\cdot\nabla\right)\boldsymbol{u}
 &= -\nabla p+\nu\nabla^2\boldsymbol{u}+\boldsymbol{f}(\boldsymbol{x}),\\
\nabla\cdot\boldsymbol{u}&=0,\qquad
\boldsymbol{f}(\boldsymbol{x})=\big(A\sin(n_f y),0\big).
\end{aligned}
\end{equation}
\qw{Energy injection at $k=n_f$ sustains a statistically stationary turbulent state.} \qw{Following the setup of \citet{cleary2025latent}, we use $Re=400$, $n_f=4$, and $A=1$.} A precursor simulation of \qw{$40$} time units from random fields \qw{supplies} the truth initial condition \qw{for the assimilation window $T=2$.} \qw{Both flows use the periodic domain and numerical method described above, and their time steps and observation intervals are listed separately in Table~\ref{tab:sim_params}.}

\qw{The velocity reconstruction error is denoted by $\epsilon(t)=\|\boldsymbol{u}(t)-\boldsymbol{u}_r(t)\|$, and is normalized by the instantaneous root-mean-square magnitude of the reference velocity fluctuations, $u_{\mathrm{rms}}(t)$, computed from both velocity components after removal of their spatial means.}

\subsection{Examples of Control-Variable Choices}
\label{sec:validation_control_variable}
From equation \eqref{eqn:new_inner_product}, the choice of control variable $\boldsymbol{s}_0$ directly determines the form of the convolution operator $\mathcal{G}$.
For instance, when the vorticity field $\omega$ is used as the control variable, the forward velocity field in Fourier space is recovered through the Biot-Savart relation,
\[
\hat{\boldsymbol{u}}(\boldsymbol{k}) = i\,\frac{\boldsymbol{k}^\perp}{k^2}\,\hat{\omega}(\boldsymbol{k}),
\qquad
\boldsymbol{k}^\perp = (-k_y,\,k_x),
\]
which implies that the kinetic energy and enstrophy spectra are connected by $E_\omega(\boldsymbol{k})=k^2 E_u(\boldsymbol{k})$.
Hence, the natural inner product between two vorticity fields $\omega_1$ and $\omega_2$ corresponds to the $\mathcal{G}$-inner product on the velocity fields
\qw{$\boldsymbol{u}_1$ and $\boldsymbol{u}_2$ with Fourier multiplier $\hat{G}(k)=1/k^{2}$,}
\begin{equation}
[\omega_1, \omega_2]
= \int_{\boldsymbol{k}} \hat{\omega}_1^{*}(\boldsymbol{k})\, \hat{\omega}_2(\boldsymbol{k})\, d\boldsymbol{k}
= \int_{\boldsymbol{k}} k^{2}\, \hat{\boldsymbol{u}}_1^{*}(\boldsymbol{k})\, \hat{\boldsymbol{u}}_2(\boldsymbol{k})\, d\boldsymbol{k}
= \langle \boldsymbol{u}_1, \boldsymbol{u}_2 \rangle_{\hat{G} = 1/k^{2}} .
\end{equation}

Similarly, the effect of using streamfunction as the control vector for the data assimilation can be assessed via $\hat{\boldsymbol{u}}(\boldsymbol{k}) = i\,\boldsymbol{k}^\perp\,\hat{\psi}(\boldsymbol{k})$. The inner product between two streamfunctions then becomes an integral of $k^{-2}\,\hat{\boldsymbol{u}}_1^*(k)\,\hat{\boldsymbol{u}}_2(k)$ in the Fourier space, showing that energy of the streamfunction scales with kinetic energy by $k^{-2}$, i.e,
\begin{equation}
\left[\psi_1, \psi_2\right] = \left\langle \boldsymbol{u}_1, \boldsymbol{u}_2 \right\rangle_{\hat{G} = k^2}.
\label{eqn:vorticity_inner_result}
\end{equation}

To assess the impact of control-variable choice on reconstruction accuracy, we perform data assimilation \qw{in forced Kolmogorov flow} using four different control vectors: streamfunction \(\psi_0\), velocity \(\boldsymbol{u}_0\), vorticity \(\omega_0\), and velocity with Sobolev regularization \qw{$\lambda = 5\times 10^{-2}$}. The value of $\lambda$ is chosen by running a short optimization (a few L-BFGS iterations) for several candidate values and selecting the one that yields the lowest objective before performing the full assimilation.

\begin{figure}[ht]
    \centering
    \includegraphics[width=\textwidth]{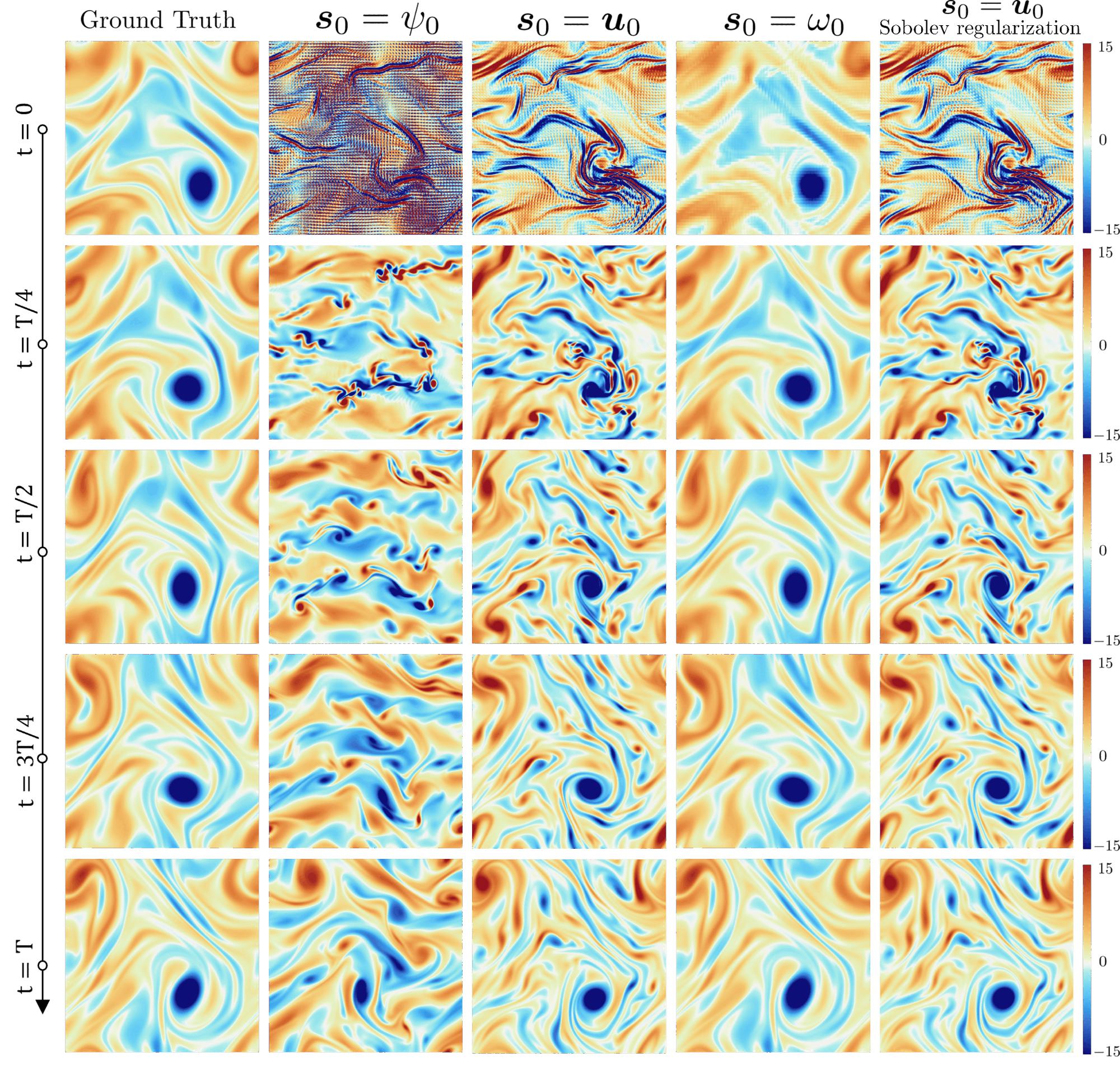}
    \caption{\qw{Reconstruction snapshots of assimilated Kolmogorov flow with different choices of control variable.} Rows show vorticity at $t=0,T/4,T/2,3T/4,T$. Columns show the truth, streamfunction control, velocity control, vorticity control, and Sobolev-regularized velocity control.}
    \label{fig:snapshots_uv}
\end{figure}

Figure~\ref{fig:snapshots_uv}  presents a time sequence of two-dimensional turbulence fields, in terms of vorticity contour, at five representative instants: $t = 0$, $T/4$, $T/2$, $3T/4$, and $T$ (rows). Each column compares the ground-truth field to model reconstructions using the control vector $\boldsymbol{s}_0$ being (i) streamfunction $\boldsymbol{s}_0 =\psi_0$, (ii) velocity components $\boldsymbol{s}_0 = \boldsymbol{u}_0$, (iii) vorticity $\boldsymbol{s}_0 = \omega_0$, and (iv) a Sobolev-regularized case with $\boldsymbol{s}_0 = \boldsymbol{u}_0$. All panels share a consistent colormap, with colorbars normalized per row to facilitate temporal comparison. Visually, \qw{except for the streamfunction case,} the reconstructions exhibit strong phase fidelity across time, with large-scale structures accurately positioned. Reconstructions based on \qw{$\omega_0$} most closely match the ground truth. When \qw{$\boldsymbol{u}_0$} is used as the control vector, the method exhibits heightened sensitivity to small-scale gradients, resulting in \qw{spurious fine-scale oscillations in the reconstructed initial condition, although the large-scale structures are recovered at later times}. \qw{Using} $\psi_0$ as the control vector leads to the emergence of \qw{strong} spurious fine-scale features \qw{and substantial errors in the large-scale evolution}, reflecting the reduced sensitivity of the optimization to such structures. The Sobolev diagnostic indicates that the method broadly preserves multiscale regularity, though \qw{at the tested penalty strength it changes the velocity-control reconstruction only slightly}.

\begin{figure}[ht]
    \centering
    \includegraphics[width=0.8\textwidth]{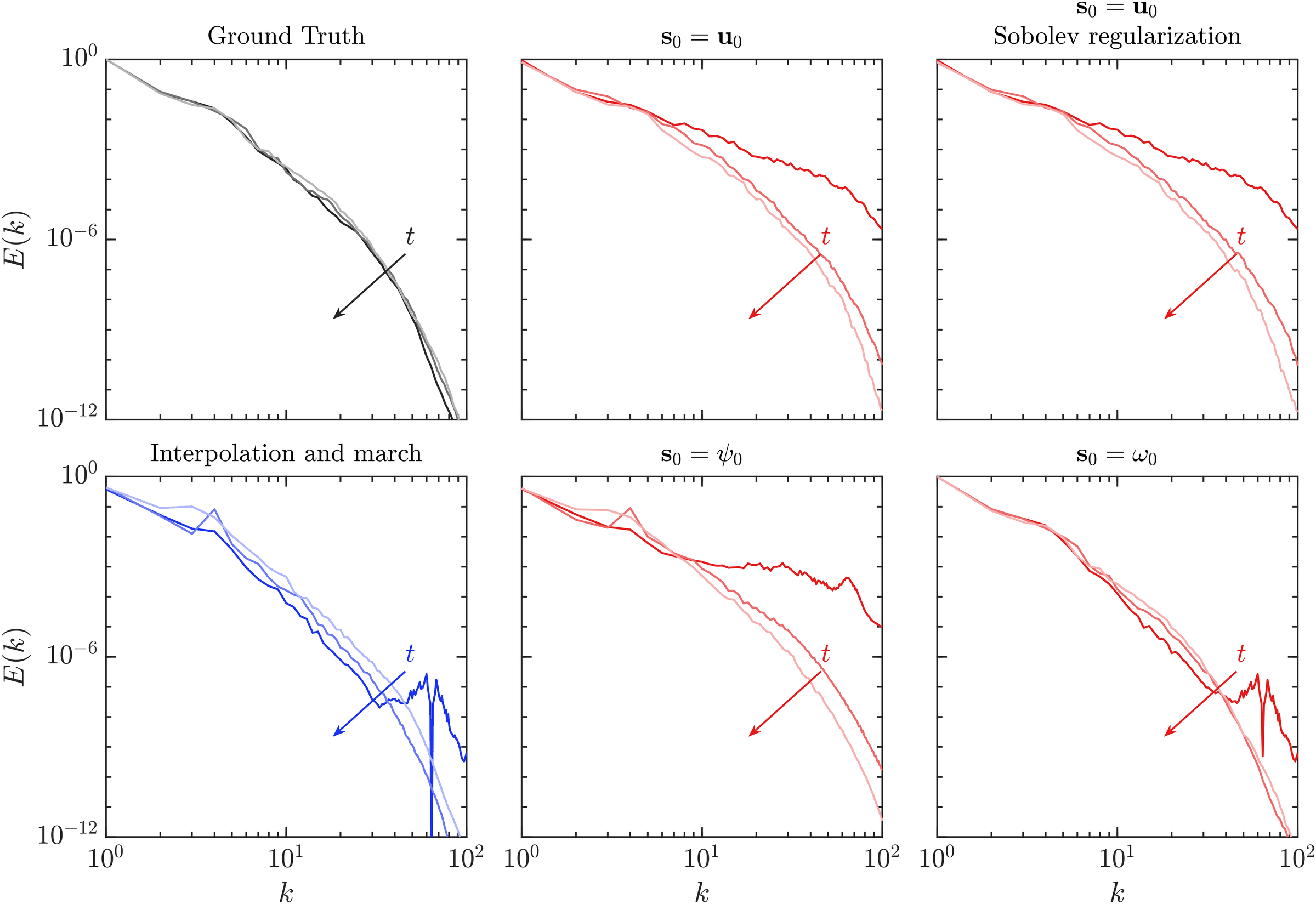}
    \caption{Energy spectra $E(k)$ for the \qw{assimilated Kolmogorov flow with different choices of control variable.} Gray curves show the truth, blue curves show the interpolated initial guess and its subsequent evolution, and red curves show the reconstructed fields. The control variables are velocity, streamfunction, vorticity, and Sobolev-regularized velocity. Lighter shades indicate later times.}
    \label{fig:cv_energyspectrum}
\end{figure}

\begin{figure}[ht]
    \centering
    \includegraphics[width=0.8\textwidth]{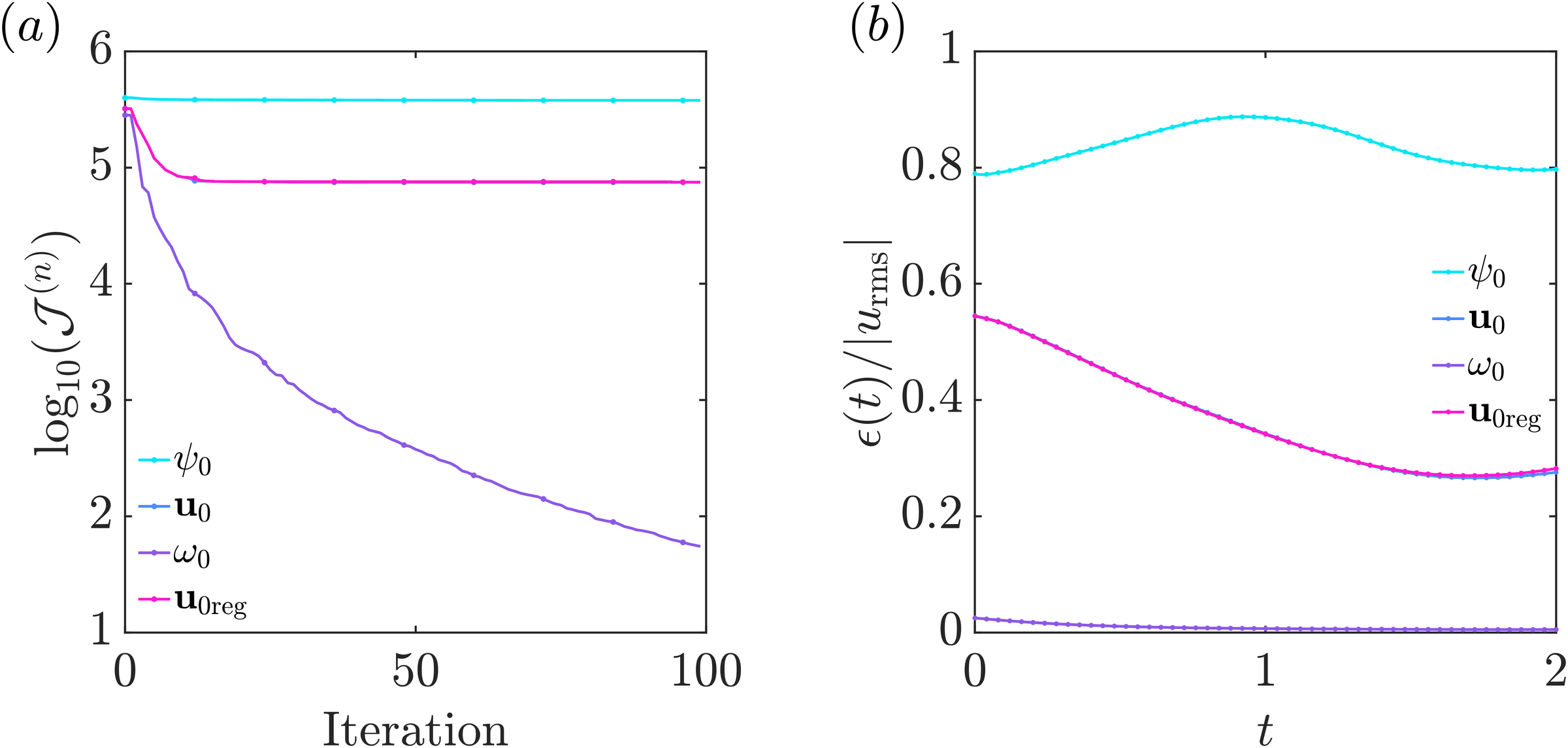}
    \caption{Optimization and reconstruction diagnostics for \qw{assimilated Kolmogorov flow with different choices of control variable.} (a) Objective} function $\mathcal{J}^{(n)}$ over L-BFGS iterations, on a logarithmic scale. (b) Velocity reconstruction error normalized by the instantaneous root-mean-square magnitude of the truth velocity fluctuations, $\epsilon(t)/u_{\mathrm{rms}}(t)$. The same colors identify each formulation in both panels: streamfunction control, velocity control, vorticity control, and Sobolev-regularized velocity control.
    \label{fig:cv_convergence}
\end{figure}

Figure \ref{fig:cv_energyspectrum} summarizes the performance of the data assimilation procedure across control variables. The energy spectra $E(k)$ for the evolution of initial fields reconstructed using different control variables are reported. Blue curves correspond to the initial interpolated state and its evolution in time. Gray curves represent ground-truth snapshots, while red curve families illustrate the evolution of reconstructions over time using different control variables $\boldsymbol{s}_0$. At the initial time $t = 0$, the reconstructions using the vorticity ($\boldsymbol{s}_0 = \omega_0$) perform comparably to the \qw{ground truth}, while all other reconstructions show different levels of mismatch, especially in the high wavenumber region. This trend reflects the intrinsic scale-dependent sensitivity of the adjoint dynamics: backward integration amplifies small-scale components more strongly, causing the optimization to favor high-wavenumber corrections unless appropriately regularized. As the flow evolves in time, all reconstructions progressively improve in terms of accuracy. In addition, the Sobolev-regularized optimization \qw{has little effect on the excess high-wavenumber energy of the velocity-control result}, while the closest recovery is the one using \qw{vorticity $\omega_0$} as the control vector.

Overall quantification of reconstruction qualities is shown in figure \ref{fig:cv_convergence}. The left panel plots the objective functional $\mathcal{J}^{(n)}$ versus iteration count $n$ for various control configurations. \qw{The vorticity} control vector \qw{$\boldsymbol{s}_0 = \omega_0$} \qw{gives} the fastest descent and lowest minimum, indicating superior optimization efficacy. \qw{Velocity control stops decreasing after about ten iterations}, while $\boldsymbol{s}_0 = \psi_0$  converges the slowest and reaches a plateau after a few iterations---consistent with its spectral difficulty.

The right panel presents the normalized root-mean-square error as a function of time and exhibits the same qualitative trends. The error decreases in time, reflecting the increasing influence of later observations relative to earlier ones. This behavior is a direct consequence of adjoint-field growth and will be examined in more detail in subsequent sections. \qw{The initial-condition error decreases from $0.545$ with velocity control to $0.025$ with vorticity control, compared with $0.789$ for streamfunction control and $0.544$ for the Sobolev-regularized case.}

This equivalence between the choice of control variable and spectral preconditioning can also be verified numerically: applying the preconditioner $\hat{G}(k)=1/k^2$ to velocity control yields a reconstruction visually indistinguishable from that obtained with vorticity control after $N=100$ L-BFGS iterations. We put a brief validation \qw{using the decaying-HIT case} in Appendix~\ref{app:cv_equivalence}.

\subsection{Construction of Spectral Filter as Preconditioner}
\label{sec:results_construction}
Motivated by this observed equivalence, we now formalize the construction of a spectral filter as a preconditioner, within which the choice of control variable, including vorticity, streamfunction, velocity, or even intermediate representations, can be treated in a unified manner.
We are particularly interested in two types of preconditioner here, 
\[
\widehat{G}_{p}(k;\alpha) = k^{-2\alpha}, \text{ and }\widehat{G}_{e}(k;\beta) = \exp\!\left(-\nu \beta\,k^{2}\right),
\]
with subscripts $p$ and $e$ indicating power law and exponential. Here we exclude the zero wavenumber mode by setting \( \widehat{G}(0) = 1 \) in the code to avoid singularities. In Fourier space, the scaled control variables for the two scenarios are respectively
\[
\widehat{\boldsymbol{s}}_p(k) = \widehat{G}_p^{-1/2}(k)\,\widehat{\boldsymbol{u}}_0(k) = k^{\alpha} \widehat{\boldsymbol{u}}_0(k),\quad
\widehat{\boldsymbol{s}}_e(k) = \widehat{G}_e^{-1/2}(k)\,\widehat{\boldsymbol{u}}_0(k) = \exp\!\left(\frac 12 \nu \beta\,k^{2}\right)\widehat{\boldsymbol{u}}_0(k).
\]
These control variables can be physically interpreted.
If we regard the vorticity and streamfunction as the order $1$ and $-1$ derivatives of a two-dimensional incompressible velocity field, the algebraic family defines the fractional derivatives of $\boldsymbol{u}_0$. Correspondingly, $\boldsymbol{u}_0$ is a fractional integral of $\boldsymbol{s}_0$ for non-integer $\alpha$.

For the preconditioner in the exponential-family, the transformed control variable uses a heat-kernel-type factor that damps high $k$ when $\beta>0$.
Equivalently, the mapping from the latent control $\boldsymbol{s}_0$ back to the physical initial condition $\boldsymbol{u}_0$ is

\begin{equation}
\widehat{\boldsymbol{u}}_{0}(k)
=\exp\!\left(-\frac{1}{2}\nu\beta k^{2}\right)\widehat{\boldsymbol{s}}_{e}(k).
\label{eq:ue_from_se}
\end{equation}

Since $\exp(\tau \Delta)$ has Fourier symbol $\exp(-\tau k^{2})$, the operator
$\exp\!\left(-\frac{1}{2}\nu\beta k^{2}\right)$ in \eqref{eq:ue_from_se}
corresponds to a diffusive (smoothing) step in the $\boldsymbol{s}_0\mapsto\boldsymbol{u}_0$ mapping, with the diffusivity $\nu$ and diffusion time $\beta/2$.
when $\beta>0$: the velocity is smoothened from the control variable $\boldsymbol{s}_0$ by $\mathcal{G}_e^{1/2}$, as in figure \ref{fig:DA_framework}.
In contrast, $\beta<0$ yields a sharpening in the $\boldsymbol{s}_0\mapsto\boldsymbol{u}_0$ mapping that further amplifies high-wavenumber modes, which is not preferable here.


\qw{These two families provide complementary forms of spectral weighting: the algebraic family gives a gradual tilt across wavenumbers, whereas the exponential family attenuates high-wavenumber components more sharply.} 

\subsection{Results with Algebraic Spectral Preconditioners}
\label{sec:results_power}
\begin{figure}[ht]
    \centering
    \includegraphics[width=\textwidth]{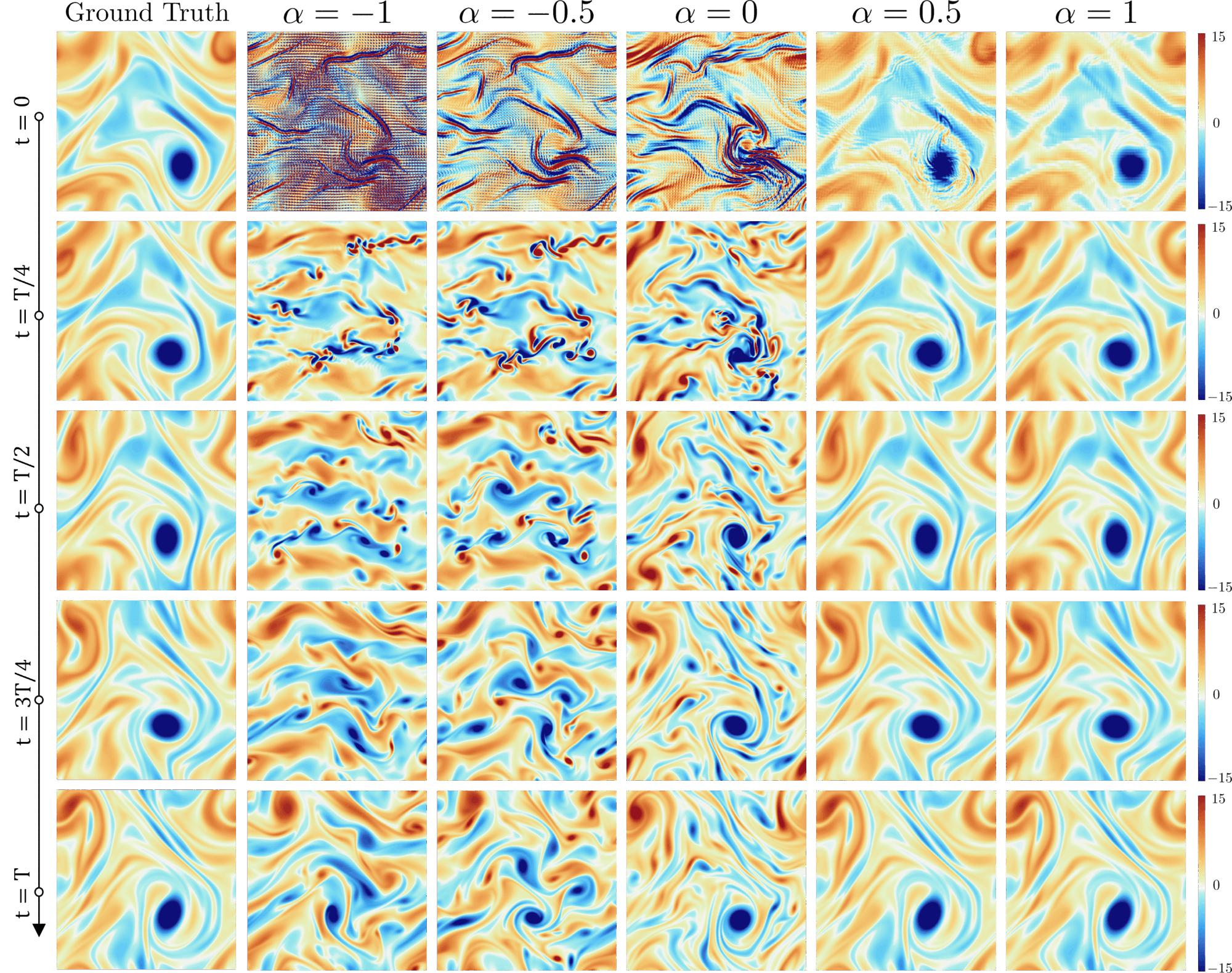}
    \caption{The reconstructed vorticity fields in \qw{forced Kolmogorov flow} for the algebraic preconditioner $\widehat{G}_p(k)=k^{-2\alpha}$ at $t=\{0,0.5,1,1.5,2\}$.}
    \label{fig:kolm_alpha_snap}
\end{figure}

\begin{figure}[ht]
    \centering
    \includegraphics[width=\textwidth]{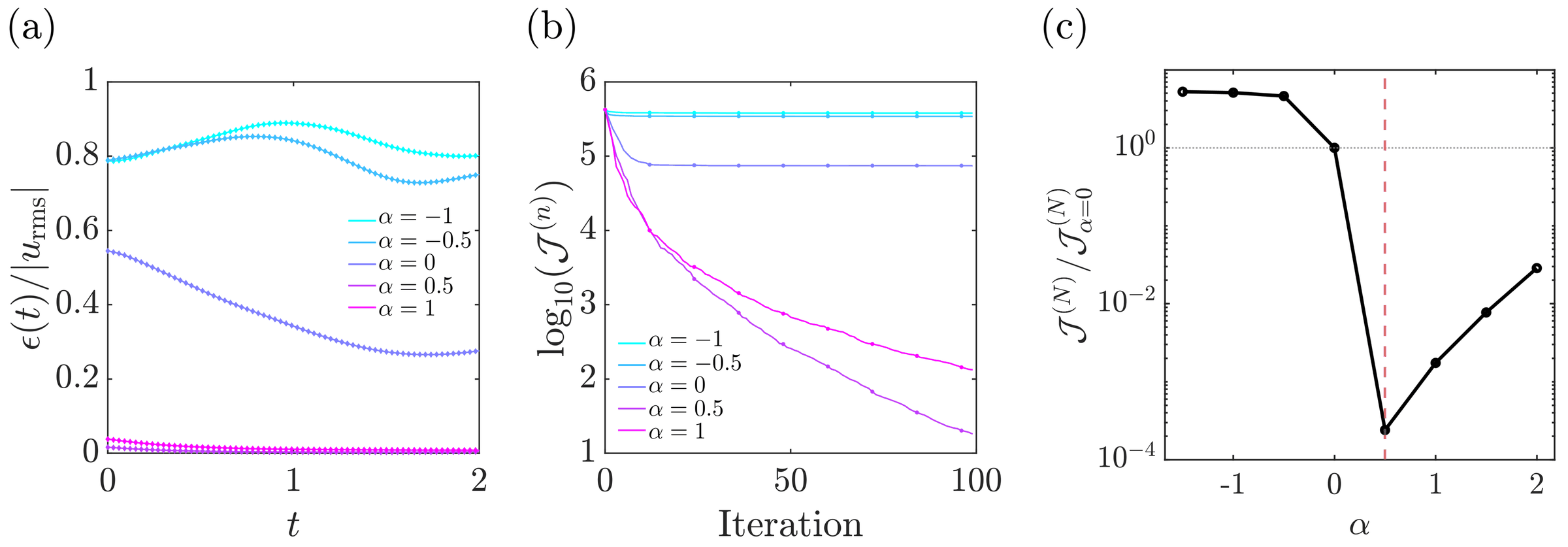}
    \caption{\qw{Forced Kolmogorov-flow} reconstruction with the algebraic preconditioner $\widehat{G}_p(k)=k^{-2\alpha}$. (a) Velocity reconstruction error $\epsilon(t)$ normalized by the instantaneous root-mean-square magnitude of the ground-truth velocity fluctuations, $u_{\mathrm{rms}}(t)$. (b) Convergence histories of the objective function $\mathcal{J}^{(n)}$ over L-BFGS iterations, plotted as $\log_{10}\mathcal{J}^{(n)}$. (c) Final objective value normalized by the unpreconditioned result, $\mathcal{J}^{(N)}/\mathcal{J}^{(N)}_{\alpha=0}$, versus $\alpha$.}
    \label{fig:kolm_alpha_conv}
\end{figure}

We first present the \qw{Kolmogorov-flow} results of the algebraic family
\[
\widehat{G}_p(k) = k^{-2\alpha}, \qquad
\widehat{\boldsymbol{s}}_p(k) = k^{\alpha}\widehat{\boldsymbol{u}}_0(k),
\]
by scanning $\alpha \in [-1.5,2]$. For each $\alpha$, the optimization starts from the same initial guess and is run with identical L-BFGS settings.
Figures \ref{fig:kolm_alpha_snap} and \ref{fig:kolm_alpha_conv} report the reconstructed snapshots and the convergence diagnostics for the algebraic family. The snapshots in figure~\ref{fig:kolm_alpha_snap} show that the moderately positive choices $\alpha\approx0.5$ and $\alpha\approx1$ both recover nearly the same large-scale vorticity structures as the truth at each displayed time. Compared with $\alpha\approx0.5$, the vorticity-based choice $\alpha\approx1$ gives a cleaner qualitative reconstruction at the initial time, suppressing the streak-like small-scale artifacts associated with adjoint amplification. Other values of $\alpha$ either leave excessive small-scale fluctuations or over-regularize the reconstructed field. The convergence diagnostics in
figure~\ref{fig:kolm_alpha_conv} further show \qw{a} U-shaped dependence of the normalized final objective value $\mathcal{J}^{(N)}$ on $\alpha$, with the minimum reached at $\alpha\approx 0.5$.
Although the vorticity-based choice ($\alpha\approx 1$) does not give the lowest final objective value, it achieves comparable quantitative accuracy \qw{and} better qualitative reconstruction across scales. In wavenumber space, moderately positive $\alpha$ improves the reconstruction across large, intermediate, and small scales by suppressing excessive high-wavenumber sensitivity while preserving the energetically important low-wavenumber components.

\qw{Over the window $T=2$, the standard adjoint yields a final objective $\mathcal{J}^{(N)}=7.5\times10^{4}$ and an initial-condition velocity error $\epsilon(t{=}0)/u_{\mathrm{rms}}=0.545$. The best algebraic kernel ($\alpha\approx0.5$) reduces these values to $17.9$ and $0.016$, respectively, a reduction of the velocity error by a factor of about $34$. The relative initial-vorticity error, $\epsilon_\omega(t{=}0)=\|\omega_{\mathrm{recon},0}-\omega_{r,0}\|_2/\|\omega_{r,0}\|_2$, decreases from $2.87$ to $0.27$. The unpreconditioned vorticity reconstruction is therefore less accurate than the trivial estimate $\omega_0\equiv0$, whose relative error is unity, whereas the preconditioned reconstruction recovers the initial field with substantially smaller error. The ensemble averages in Section~\ref{sec:sensitivity} confirm that this improvement is not specific to the present realization.}

\subsection{Results with Exponential (Diffusion-Like) Spectral Preconditioner}
\label{sec:results_exponential}
\begin{figure}[ht]
    \centering
    \includegraphics[width=\textwidth]{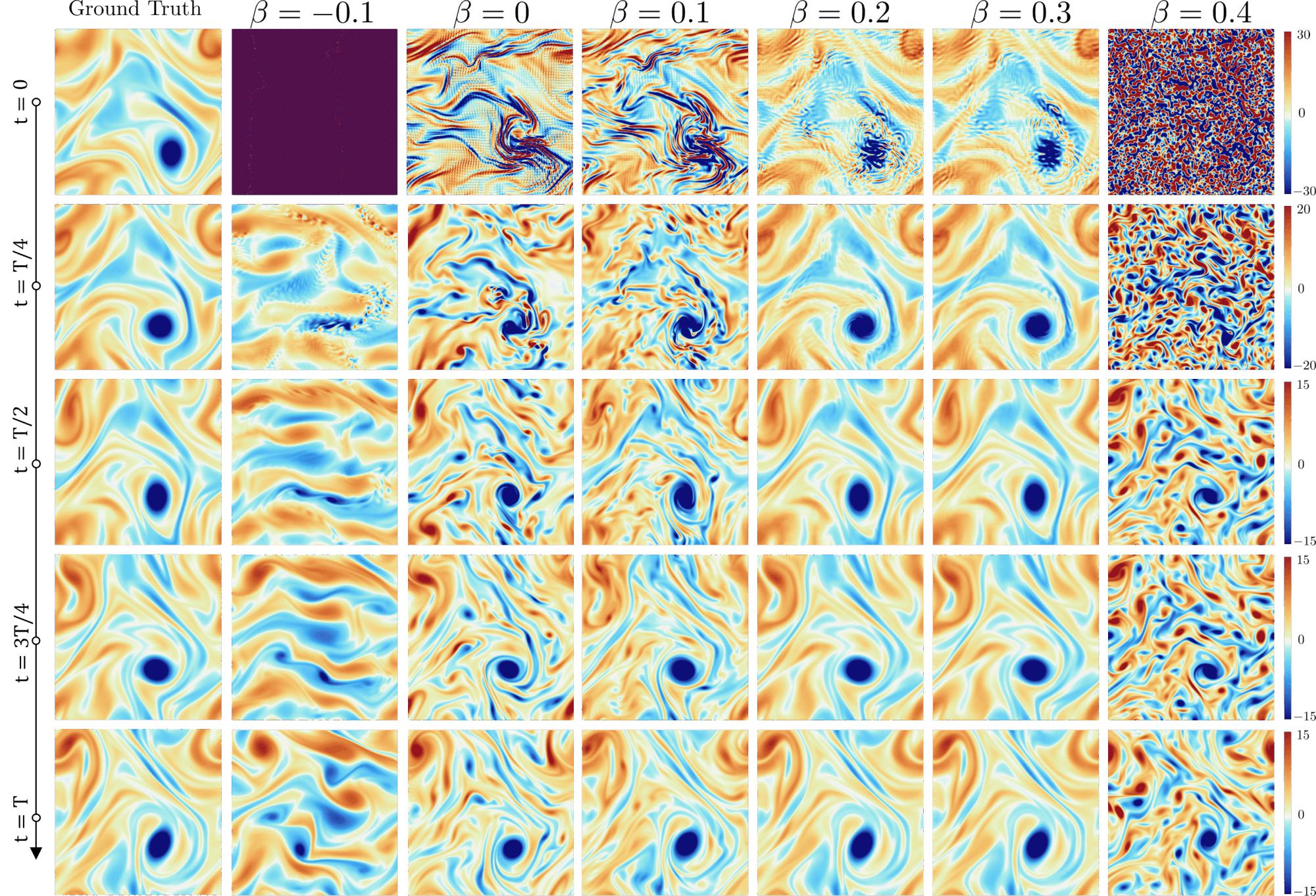}
    \caption{True and reconstructed vorticity fields in \qw{forced Kolmogorov flow} for the exponential preconditioner $\widehat{G}_e(k)=\exp(-\nu\beta k^2)$ at $t=\{0,0.5,1,1.5,2\}$.}
    \label{fig:kolm_beta_snap}
\end{figure}

\begin{figure}[ht]
    \centering
    \includegraphics[width=\textwidth]{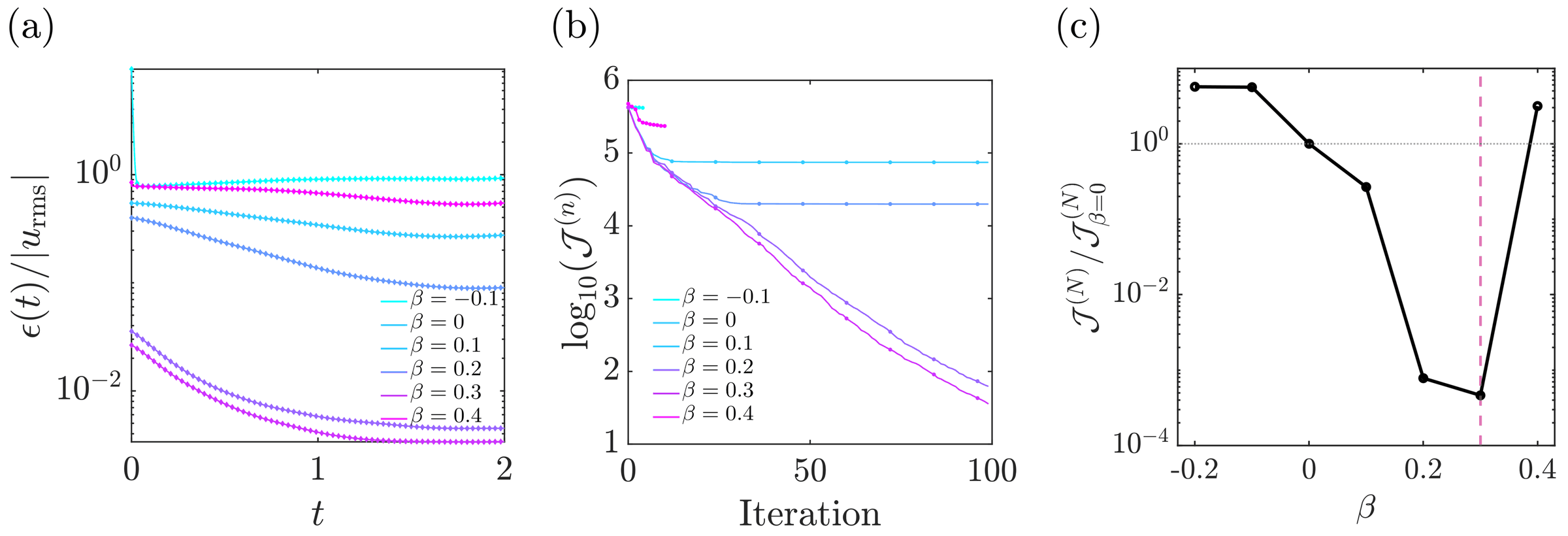}
    \caption{\qw{Forced Kolmogorov-flow} reconstruction with the exponential preconditioner $\widehat{G}_e(k)=\exp(-\nu\beta k^2)$. (a) Velocity reconstruction error $\epsilon(t)$ normalized by the instantaneous root-mean-square magnitude of the ground-truth velocity fluctuations, $u_{\mathrm{rms}}(t)$. (b) Convergence histories of the objective function $\mathcal{J}^{(n)}$ over L-BFGS iterations, plotted as $\log_{10}\mathcal{J}^{(n)}$. (c) Final objective value normalized by the unpreconditioned result, $\mathcal{J}^{(N)}/\mathcal{J}^{(N)}_{\beta=0}$, versus $\beta$.}
    \label{fig:kolm_beta_conv}
\end{figure}

The exponential preconditioner $\widehat{G}_e(k)=\exp(-\nu\beta k^2)$ is applied to the same Kolmogorov configuration and scanned over $\beta\in[\qw{-0.2},\qw{0.4}]$.
Figure~\ref{fig:kolm_beta_snap} shows that moderate positive values of $\beta$ recover the dominant large-scale vorticity structures at each displayed time while suppressing the excessive small-scale fluctuations present in the weakly filtered or unpreconditioned reconstructions. In contrast, overly weak filtering leaves visible high-wavenumber artifacts, whereas overly strong filtering distorts the reconstructed field and loses dynamically relevant structure. 
The convergence behavior in figure~\ref{fig:kolm_beta_conv}(a-b) shows that only a narrow range of parameters achieves a stable L-BFGS descent, whereas both under- and over-filtering lead to a poor convergence.
In wavenumber space, the same trend is observed: moderate exponential filtering gives the most accurate reconstruction across scales. Intermediate values such as $\beta=0.3$ reduce high-wavenumber residuals without excessively damping dynamically relevant spectral content, whereas weaker filtering leaves small-scale errors and stronger filtering over-smooths the reconstruction.
Figure~\ref{fig:kolm_beta_conv}(c) again exhibits a qualitative U-shape in the final objective $\mathcal{J}^{(N)}$, with the minimum reached at $\beta\approx 0.3$ and a rapid deterioration on either side.
These results indicate that the underlying action of the exponential preconditioner \qw{parallels that of the algebraic family}. By selectively attenuating incoherent high-wavenumber adjoint components, it improves both optimization stability and reconstruction quality. \qw{Comparing figures~\ref{fig:kolm_alpha_conv} and \ref{fig:kolm_beta_conv}, the best algebraic kernel achieves the lower final objective and initial-condition velocity error in this forced-flow case.} However, compared with \qw{the decaying HIT case in Section~\ref{sec:results_hit}}, the range of acceptable parameter values is noticeably narrower, suggesting that the forced Kolmogorov flow is less tolerant to over- or under-filtering.

\subsection{\qw{Preconditioned Adjoint Data Assimilation in 2D Decaying Homogeneous Isotropic Turbulence}}
\label{sec:results_hit}
\qw{We next apply the same two preconditioner families to two-dimensional decaying homogeneous isotropic turbulence (HIT), whose setup is described at the beginning of Section~\ref{sec:results}.}
We first present the results of the algebraic family \qw{$\widehat{G}_p(k) = k^{-2\alpha}$} by scanning $\alpha \in [-2,2]$ with identical adjoint settings.

\begin{figure}[ht]
    \centering
    \includegraphics[width=\textwidth]{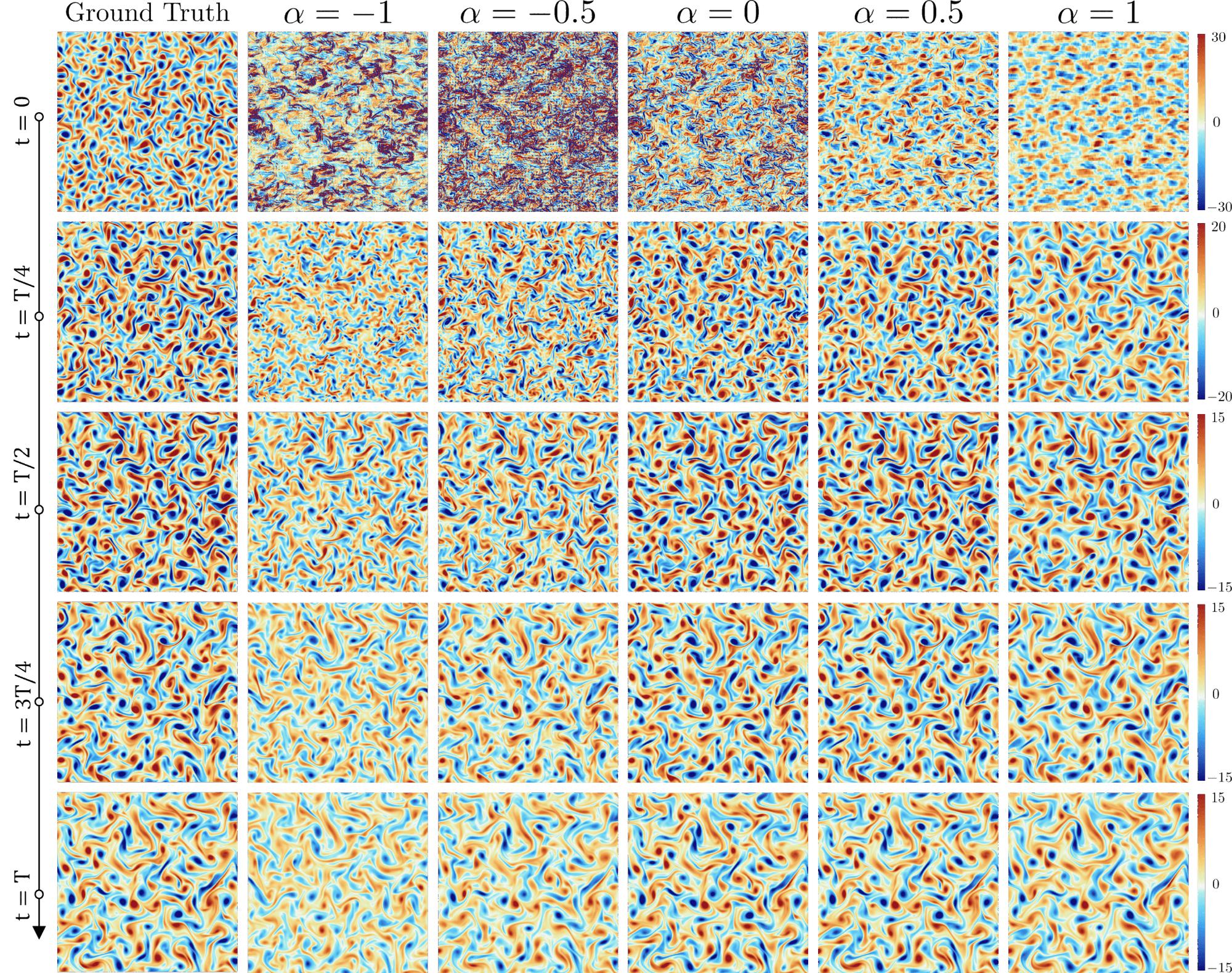}
    \caption{\qw{Decaying-HIT} vorticity reconstruction for the algebraic preconditioner $\widehat{G}_p(k)=k^{-2\alpha}$ at $t=\{0,0.5,1,1.5,2\}$.}
    \label{fig:alpha_snapshots}
\end{figure}

\begin{figure}[ht]
    \centering
    \includegraphics[width=\textwidth]{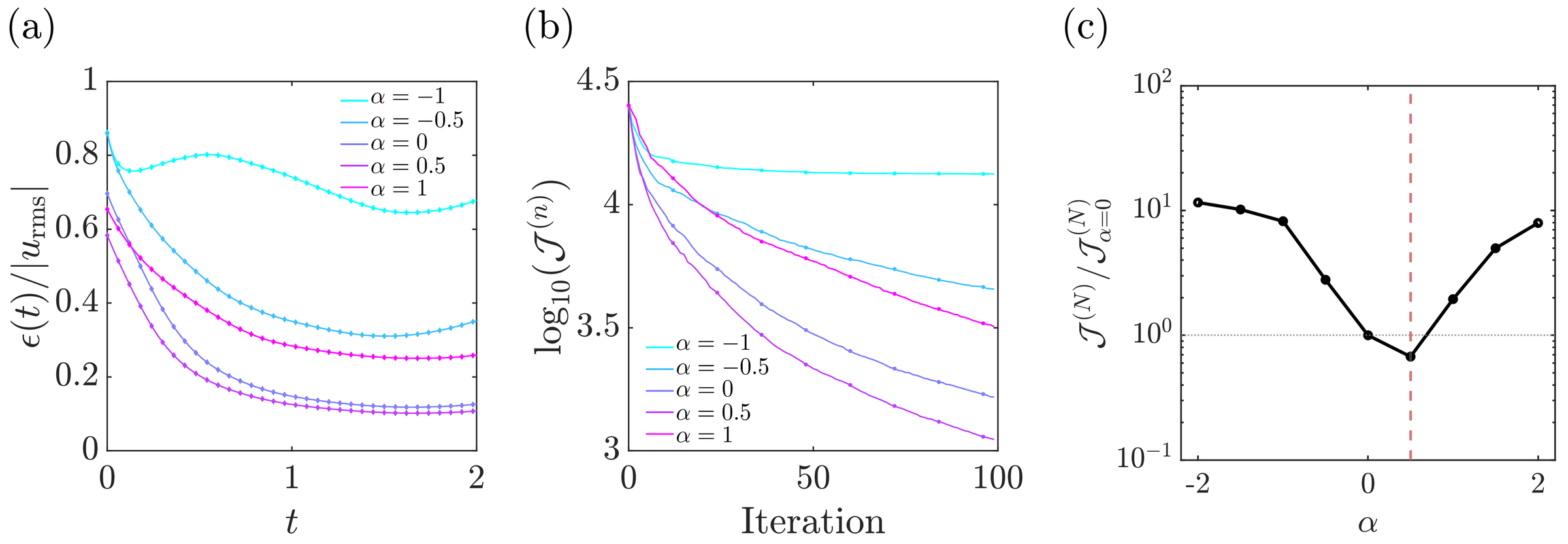}
    \caption{\qw{Decaying HIT} reconstruction quality for the algebraic preconditioner $\widehat{G}_p(k)=k^{-2\alpha}$. \qw{(a) Velocity reconstruction error $\epsilon(t)/u_{\mathrm{rms}}(t)$ as a function of time.} (b) Convergence histories of \qw{$\log_{10}\mathcal{J}^{(n)}$} over L-BFGS iterations for \qw{representative} values of \qw{$\alpha$.} \qw{(c) Final objective value normalized by the unpreconditioned result, $\mathcal{J}^{(N)}/\mathcal{J}^{(N)}_{\alpha=0}$, versus $\alpha$.}}
    \label{fig:alpha_convergence}
\end{figure}

Figure \ref {fig:alpha_snapshots} illustrates the reconstruction quality for different values of the spectral exponent $\alpha$. For moderate $|\alpha|$, the reconstructed vorticity fields capture both large-scale coherence and small-scale enstrophy with minimal spurious oscillations. \qw{As $\alpha$ approaches $\pm 1$,} positive $\alpha$ resolves finer structures but increases stiffness, whereas negative $\alpha$ yields smoother initial fields that fit large scales well but tend to under-represent small-scale vortices.

The corresponding optimization behavior is shown in figure \ref{fig:alpha_convergence}.
In all cases, the error \qw{in panel $(a)$} decreases in forward time. \qw{In addition to the influence of later observations discussed in Section~\ref{sec:validation_control_variable},} two-dimensional decaying homogeneous isotropic turbulence exhibits progressive spectral smoothing due to viscous dissipation, which reduces discrepancies as the flow evolves.
Panel $(c)$ shows the final objective value $\mathcal{J}^{(N)}$ as a function of $\alpha$. Note that $\mathcal{J}$ decreases as $\alpha$ moves from negative to mildly positive values, reaching its minimum near $\alpha \approx 0.5$. This choice slightly favors enstrophy-bearing small scales, strengthening high-$k$ corrections enough to compensate for under-resolved gradients induced by coarse observations, while keeping the adjoint dynamics and the L-BFGS iterations numerically stable. For excessively large $|\alpha|$, the L-BFGS step size often collapses, and the optimization terminates early, indicating severe ill-conditioning. Overall, the observed U-shaped dependence of $\mathcal{J}^{(N)}$ on $\alpha$ quantifies the trade-off introduced by spectral preconditioning: small $|\alpha|$ provides the best balance between stability and accuracy, while very large $|\alpha|$ either over-amplifies high-$k$ sensitivity or over-damps small-scale corrections.

\begin{figure}[ht]
    \centering
    \includegraphics[width=\textwidth]{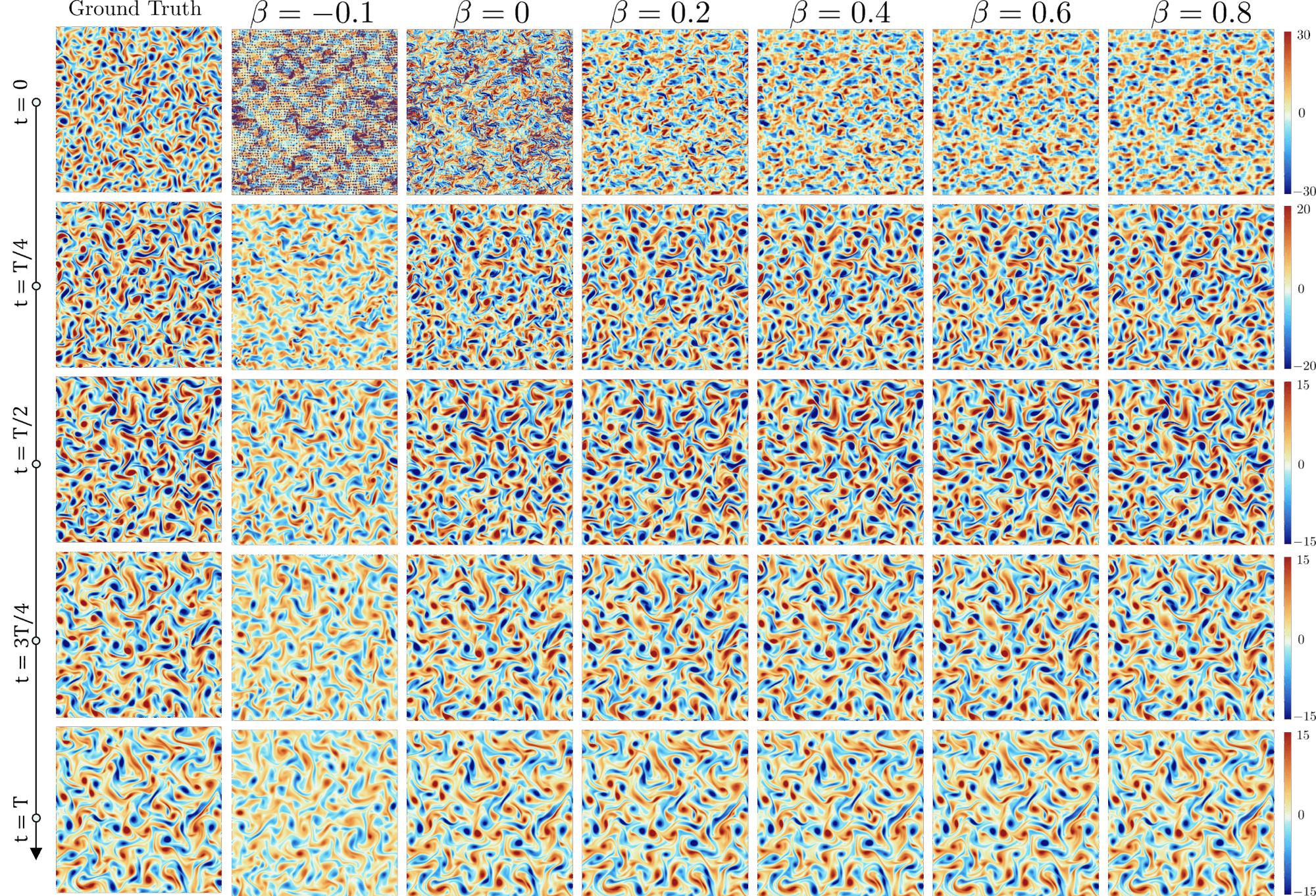}
    \caption{\qw{Decaying-HIT} vorticity reconstruction for the exponential preconditioner $\widehat{G}_e(k)=\exp(-\nu\beta k^2)$ at $t=\{0,0.5,1,1.5,2\}$.}
    \label{fig:beta_snapshots}
\end{figure}

For the exponential family, we systematically scan $\beta$ over the interval $[-0.1,0.8]$ with a uniform step size with $\Delta\beta=0.05$, starting from the same initial guess and using identical L-BFGS settings as in the algebraic family.
Figure \ref {fig:beta_snapshots} illustrates the effect of $\beta$ on the reconstruction quality, with $\widehat{G}_{e}(k) = \exp(-\nu \beta\,k^{2})$. As $\beta$ increases to moderate positive values, the reconstructed initial fields become smoother and more physically plausible: small-scale noise in the adjoint gradient is efficiently damped while large-scale structures are preserved.

\begin{figure}[ht]
    \centering
    \includegraphics[width=\textwidth]{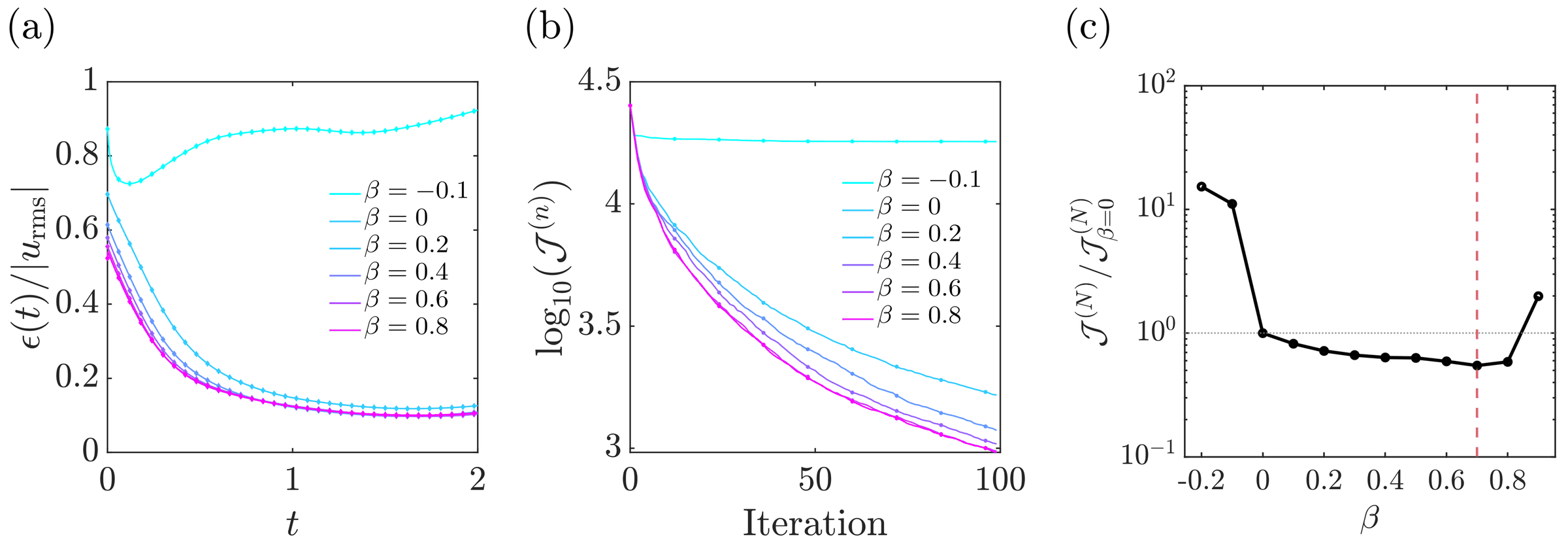}
    \caption{\qw{Decaying-HIT} reconstruction quality for the exponential preconditioner $\widehat{G}_e(k)=\exp(-\nu\beta k^2)$. (a) Velocity reconstruction error $\epsilon(t)/u_{\mathrm{rms}}(t)$ as a function of time after $100$ L-BFGS iterations. (b) Convergence histories of $\log_{10}\mathcal{J}^{(n)}$ for representative values of $\beta$. (c) Final objective value normalized by the unpreconditioned result, $\mathcal{J}^{(N)}/\mathcal{J}^{(N)}_{\beta=0}$, versus $\beta$.}
    \label{fig:beta_convergence}
\end{figure}

The corresponding optimization behavior is shown in figure \ref{fig:beta_convergence}.
The temporal evolution of the velocity mismatch exhibits trends similar to those observed with the algebraic preconditioner, as illustrated in panel (a). However, the exponential family yields the largest improvement in the reconstruction of the initial condition. Compared with the standard (unweighted) adjoint formulation, the best exponential kernel ($\beta\approx0.7$) reduces the initial-condition error $\epsilon(t{=}0)/u_{\mathrm{rms}}$ by $22\%$, from $0.70$ to $0.54$.
In panel (b), it is clear that using a negative $\beta$ leads to stagnation of the optimization, with the cost function failing to decrease after a few iterations.
The final loss in panel $(c)$ decreases as $\beta$ increases to moderate positive values. The best overall performance is obtained at an intermediate value, around $\beta \approx 0.7$, where the preconditioner achieves a favorable balance between selective high-$k$ damping and retention of dynamically relevant large-scale structure. For very large values of $\beta$ (e.g., $\beta \approx 1.0$, which lies outside the range shown in the figure), the gradient becomes excessively smoothed, effectively removing almost all high‑wavenumber corrections. As a result, the optimization loses its ability to recover fine‑scale structures. In this regime, the problem also becomes numerically stiff, and the L‑BFGS step size may collapse, causing the optimization to terminate prematurely and yielding a larger final loss. \qw{Thus, the favorable degree of filtering differs from that in Kolmogorov flow, where the exponential optimum is near $\beta\approx0.3$.}

\begin{figure}[ht]
    \centering
    \includegraphics[width=0.8\textwidth]{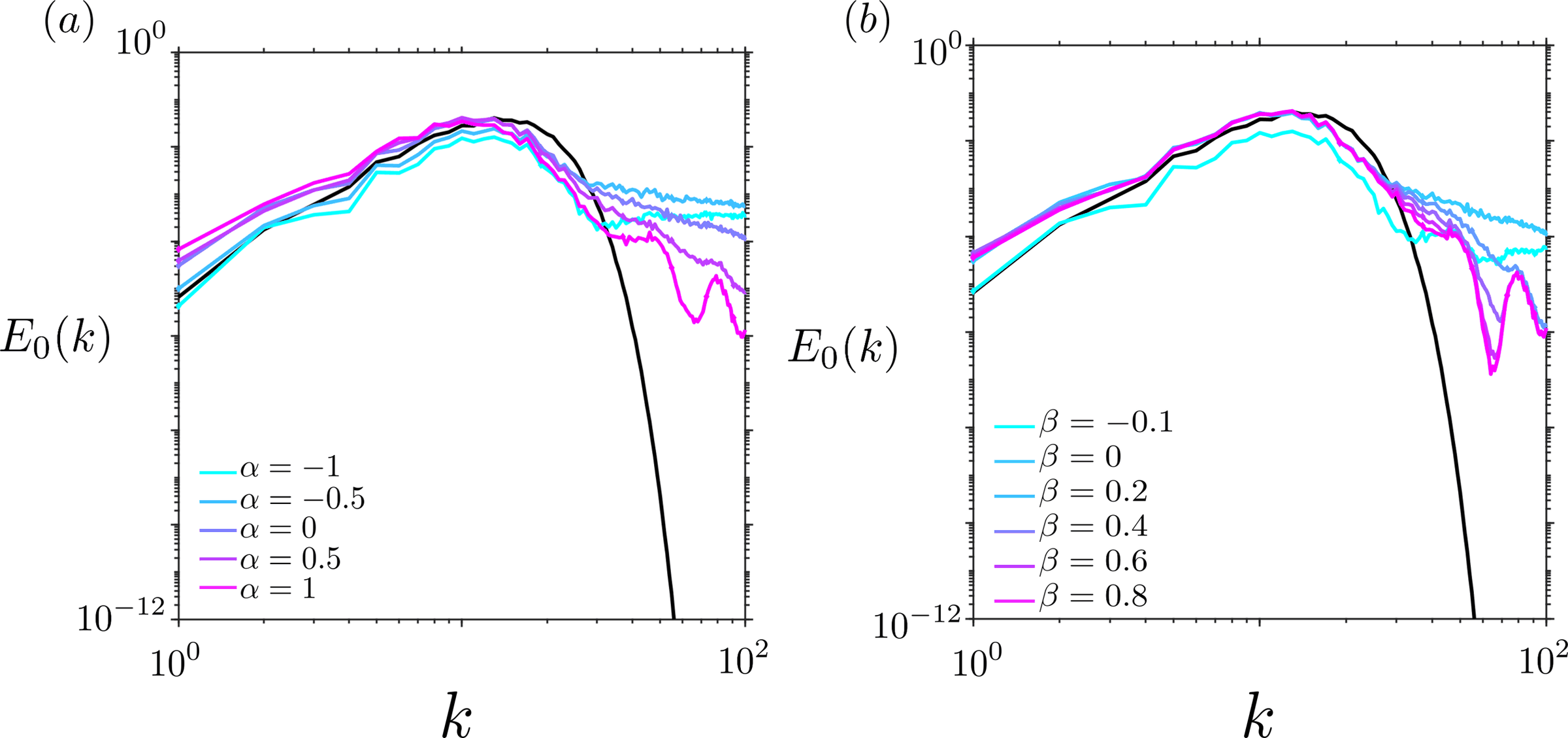}
    \caption{Energy spectrum $E_0(k)$ of the reconstructed initial condition for \qw{decaying HIT}. The left panel shows the algebraic family $\widehat{G}_p(k)=k^{-2\alpha}$, and the right panel shows the exponential family $\widehat{G}_e(k)=\exp(-\nu\beta k^2)$. The black curve in each panel is the ground-truth spectrum.}
    \label{fig:hit_alpha_beta_spec}
\end{figure}

\qw{Figure~\ref{fig:hit_alpha_beta_spec} compares the reconstructed initial energy spectra for the two families. Positive algebraic and exponential weights both reduce the excess high-wavenumber energy relative to the unpreconditioned reconstruction. The exponential family provides stronger attenuation of the high-wavenumber tail for the illustrated positive parameter values.}

\qw{As a result, the preferred kernel family differs between the two flows: the exponential family most closely reproduces the target spectrum in decaying HIT, whereas the Kolmogorov-flow results favor the algebraic family.} This reversal is consistent with the different spectral characteristics of the two flows. The decaying HIT spectrum develops a pronounced viscous-range decay, making the heat-kernel weighting $\exp(-\nu\beta k^2)$ naturally compatible with the underlying spectral structure. In contrast, the statistically stationary Kolmogorov flow is dominated by inertial-range power-law scaling, for which the algebraic weighting $k^{-2\alpha}$ provides a closer spectral match. \qw{This spectral interpretation explains the preferred kernel family, while the much larger improvement in Kolmogorov flow is associated with a wider separation between energy-containing and preconditioned scales.}

\qw{The scale separation can be quantified from the energy spectra. In Kolmogorov flow the spectrum peaks at $k=1$ and $99\%$ of the energy lies below $k=5$, whereas in decaying HIT the peak is at $k=11$ and the corresponding threshold is $k=34$. Taking the exponential kernel as a common diagnostic, its $e^{-1}$ attenuation wavenumber $k_e=(\nu\beta)^{-1/2}$ is nearly the same in both flows ($k_e\approx37$ for $\beta=0.3$ in Kolmogorov flow and $k_e\approx38$ for $\beta\approx0.7$ in HIT), and in both flows the standard-adjoint error lies predominantly above the energy-containing band ($96\%$ and $76\%$, respectively). The energy-containing and attenuated bands are therefore separated by almost a decade in Kolmogorov flow ($k_e/k_{99}\approx7$) but nearly coincide in HIT ($k_e/k_{99}\approx1.1$), so that high-wavenumber contamination can be removed at much lower cost to the energetic scales in the forced case. This interpretation is consistent with \citet{inubushi2026synchronisation}, who identify the forcing scale as the essential observation resolution in two-dimensional Kolmogorov flow. Predicting the optimal kernel systematically from the flow spectrum remains a topic for future work.}

\subsection{\qw{Sensitivity to Observation Resolution and Assimilation Window}}
\label{sec:sensitivity}

\qw{We finally examine whether the benefit of} preconditioning persists when the observation resolution \qw{and the assimilation window are varied, for both flows. The preconditioners are held fixed near their optimal values while one parameter is varied at a time.} \qw{We vary} the spatial observation gap $\Delta I$ and the \qw{observation interval $\Delta t_m$ at fixed $T=2$, and the window length $T$ at the baseline $\Delta I$ and $\Delta t_m$, so that the number of observation times grows in proportion to $T$.} Each case is run for $100$ L-BFGS iterations and averaged over $10$ independent realizations. \qw{The optimal $\alpha$ and $\beta$ are expected to depend on $T$, so the preconditioned results for $T\neq2$ should be regarded as conservative. The tested $\Delta I$ and $\Delta t_m$ are} comparable to the Taylor length and time scales of the flows, consistent with the scale-based observation-resolution criterion of \citet{wang2021state}, who showed in turbulent channel-flow state estimation that the critical data resolution scales with the Taylor microscale.

\begin{figure}[ht]
    \centering
    \includegraphics[width=\textwidth]{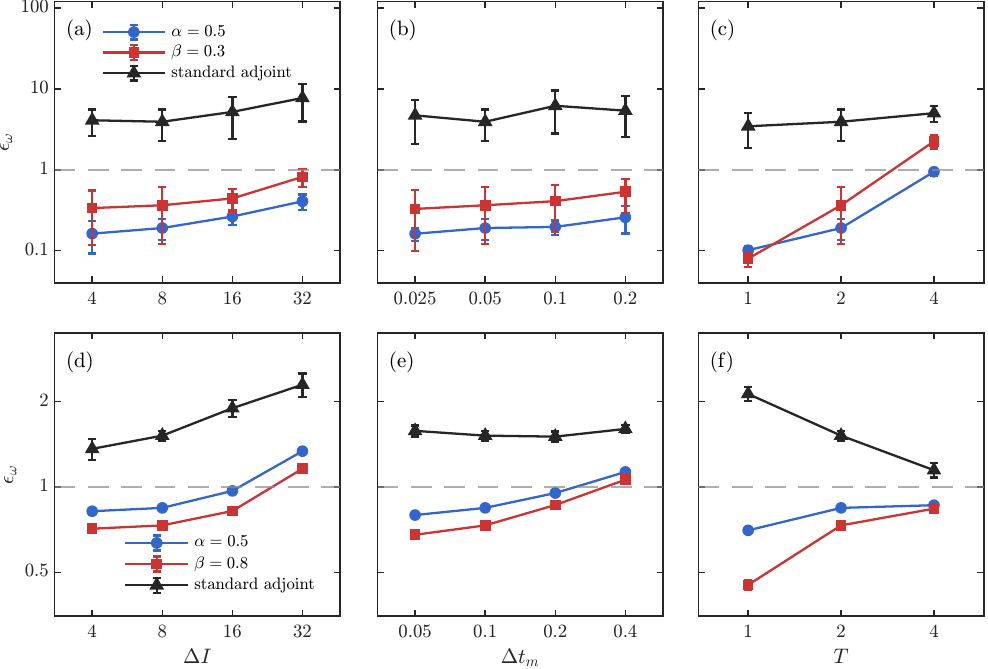}
    \caption{Sensitivity of the relative initial-vorticity error $\epsilon_\omega(t{=}0)=\|\omega_{\mathrm{recon},0}-\omega_{r,0}\|_2/\|\omega_{r,0}\|_2$ to \qw{\textbf{(a,d)}} the spatial observation gap $\Delta I$, \qw{\textbf{(b,e)} the observation interval $\Delta t_m$, and \textbf{(c,f)} the assimilation window $T$, for forced Kolmogorov flow (top row, $\alpha=0.5$, $\beta=0.3$) and decaying HIT (bottom row, $\alpha=0.5$, $\beta=0.8$).} Each point is the mean over $10$ \qw{independent realizations;} error bars denote one standard deviation.}
    \label{fig:sensitivity}
\end{figure}

\qw{Figure~\ref{fig:sensitivity} shows $\epsilon_\omega(t=0)$ for both flows. The preconditioned adjoint data assimilation is consistently more accurate than the standard adjoint across all observation settings and both flows. The poor performance of the standard adjoint stays nearly the same with more observation times, as shown in $(b)$ and $(e)$. Likewise, even the densest spatial sampling does not reduce the error below unity. Its failure therefore reflects poor gradient conditioning rather than a lack of observational information. In addition, the dependence on the assimilation window differs across methods and flows, as shown in $(c)$ and $(f)$.}

\qw{The decrease of the standard-adjoint error with $T$ in decaying HIT, but not in Kolmogorov flow, reflects the decay of the flow itself. Separating the HIT error into wavenumber bands shows that the large-scale contribution grows with $T$ for all three methods, whereas the spurious small-scale content of the standard adjoint decreases by a factor of about six from $T=1$ to $T=4$. Between $t=0$ and $t=4$ the enstrophy of the truth drops by more than an order of magnitude, so the late-time observations become insensitive to small-scale features of the initial condition and no longer drive their reconstruction. This implicit regularization is supplied by the flow rather than by the assimilation method. The preconditioned formulations already suppress this band explicitly and therefore gain little from the additional decay.}

\qw{This comparison also shows that adding more observations within a fixed window is not equivalent to extending the window. At $T=2$, increasing the number of observation times to $40$ with $\Delta t_m=0.05$ leaves the standard-adjoint error at $1.58$. Spreading the same $40$ observations over $T=4$ reduces the error to $1.15$.}

\section{Statistical Analysis of Adjoint Sensitivity}
\label{sec:stats_adjoint}
To understand the gradient information used in the optimization and its scale dependence, we analyze the spatiotemporal evolution of adjoint fields using an ensemble-based impulse response method for decaying HIT.
We refer interested readers to a similar analysis repeated for statistically stationary Kolmogorov flow to Appendix~\ref{app:dod_kolmogorov}.
We consider a terminal point perturbation in the $u$ component at $t=T$ and propagate it backward using the adjoint equations, i.e., evaluating the adjoint Green's function.
Similar to our previous research \citep{wang_wang_zaki_2022, wang2025domain}, this adjoint Green's function effectively defines the spatiotemporal domain of dependence for a later measurement (in this case, the $u$ component at a single point).
Since the adjoint dynamics depend on the underlying forward realization of homogeneous isotropic turbulence, we generate an ensemble of forward fields sharing the same initial energy spectrum $E_{0,r}$, but with random phases, and evolve these fields forward in time. Each realization is evolved forward in time, and the corresponding adjoint field is then integrated backward from the same terminal measurement kernel. This procedure yields an ensemble of adjoint fields, which enables statistical characterization of their scale-dependent structures.

\begin{figure}[h]
    \centering
\includegraphics[width=\textwidth]{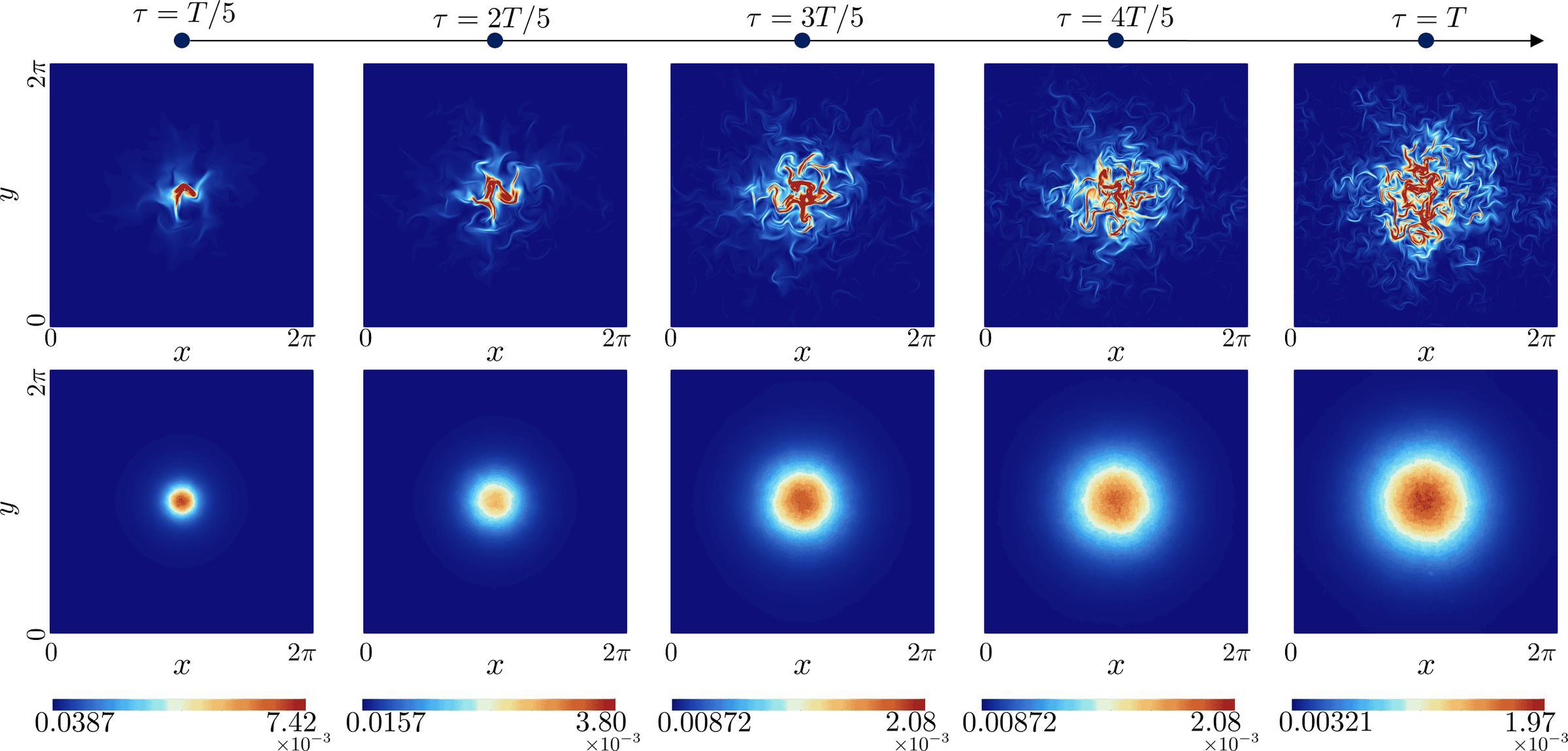}
    \caption{Top row: Randomly selected sample from the ensemble. Bottom row: Ensemble-averaged domain of dependence for the adjoint field. Colored contours show the magnitude of the adjoint velocity vector, $|\boldsymbol{u}^{\dagger}|$.}
    \label{fig:DoD}
\end{figure}

Figure \ref {fig:DoD} visualizes the adjoint fields in backward time $\tau = T-t$, in terms of the magnitude of adjoint velocity fields, $|\boldsymbol{u}^{\dagger}|$. The adjoint fields characterize the region at an earlier time $t$ that influences the observation at the terminal time $t=T$.
The top row of figure \ref{fig:DoD} displays a single realization from the ensemble.
The adjoint field is characterized by fine-scale, filamentary structures distributed throughout the domain, reflecting the chaotic nature of the flow, in which infinitesimal perturbations undergo exponential amplification when evolved backward in time. This behavior is closely connected to the Lyapunov spectrum of the system, as discussed in \citet{nikitin2018characteristics,wang_wang_zaki_2022}. The filament thickness in the adjoint vorticity field is extremely small, consistent with the Kolmogorov length scale governing the smallest dynamically active structures in the flow \citep{du2025vorticity}.
In contrast, the bottom row shows the ensemble-averaged adjoint fields averaged over 3000 realizations of turbulence for the HIT case. Here, the chaotic fluctuations cancel out, revealing a smooth core domain of dependence. This distinct difference suggests that while the instantaneous sensitivity is dominated by chaotic noise, the underlying physical connection between the terminal observation and the initial state remains deterministic but is obscured by small-scale instabilities --- a typical backward chaotic behavior \citep{eyink2004ruelle}.

We also analyze the mean adjoint fields in Fourier space, with shell-averaged energy content, $\|\overline{\hat{\boldsymbol{u}}^{\dagger}}(k,\tau)\|^{2}$, and the ensemble-averaged adjoint energy,
$\overline{\|\hat{\boldsymbol{u}}^{\dagger}(k,\tau)\|^{2}}$, obtained by averaging the shell-wise energy over realizations. The former quantifies the large-scale structures, the reproducible component of adjoint sensitivity at scale $k$, whereas the latter represents the total adjoint energy, including small-scale structure fluctuations.

\begin{figure}
\centering
\includegraphics[width=0.55\textwidth]{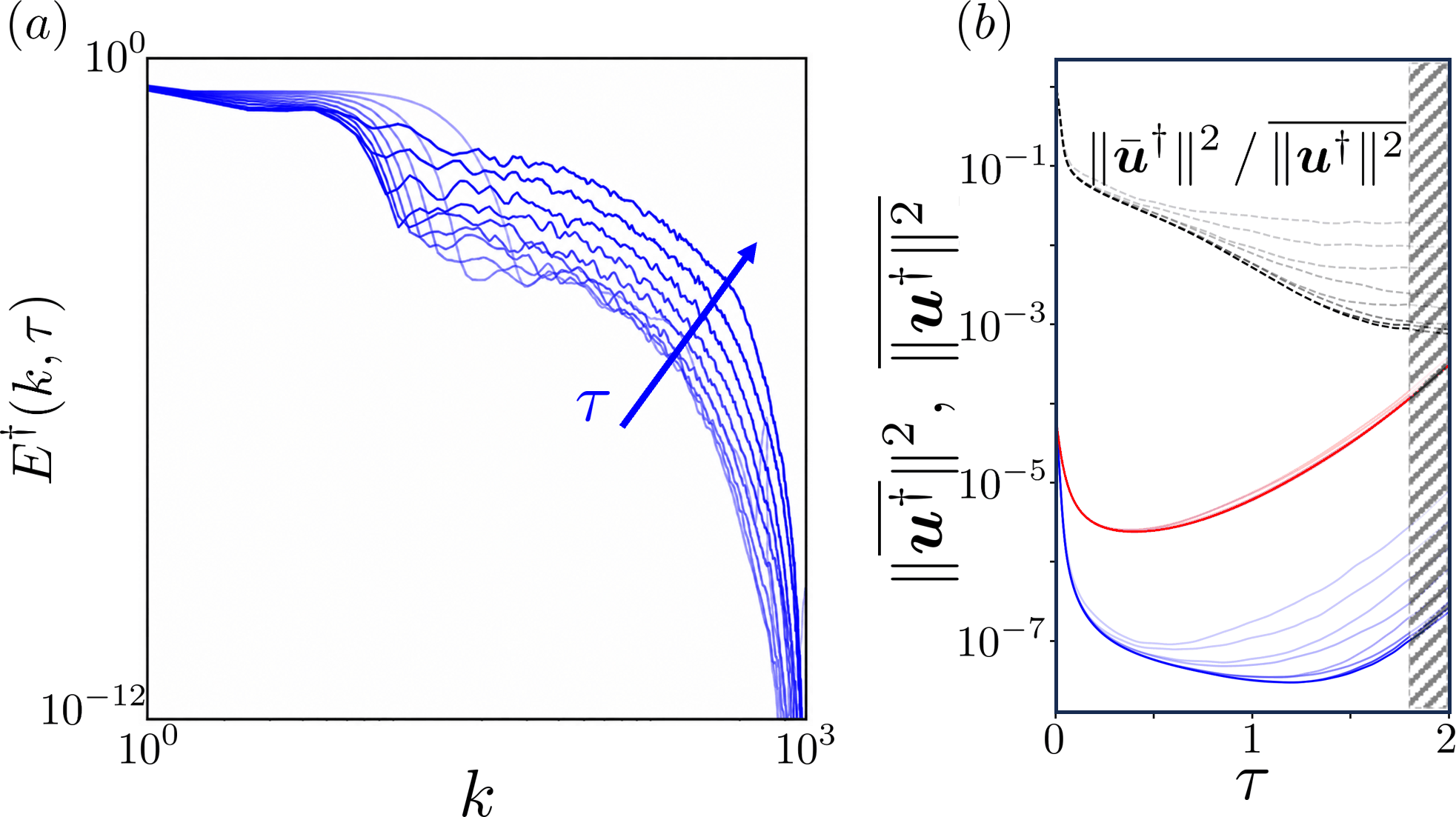}
\caption{Ensemble adjoint statistics from the domain-of-dependence calculation for two-dimensional decaying isotropic turbulence ($Re=1000$, $T=2$) , shown as functions of the backward time $\tau=T-t$. (a) Shell-averaged adjoint energy spectra $E^{\dagger}(k,\tau)$, with darker shades indicating later backward times. (b) Energy of averaged adjoint $\|\overline{\boldsymbol{u}^{\dagger}}\|^2$, average of the energy of adjoint fields $\overline{\|\boldsymbol{u}^{\dagger}\|^2}$, and their ratio. From lighter to darker curves, the ensemble subset sizes are $\{50,100,200,400,1000,3000\}$. The hatched band in (b) marks the spin-up window excluded from the analysis.}
\label{fig:dod_spectrum}
\end{figure}

Figure \ref{fig:dod_spectrum}(a) shows the shell-averaged adjoint spectra for decaying HIT. 
The adjoint energy in both flows evolves non-uniformly across wavenumbers during backward integration, with high-wavenumber components amplifying more rapidly than the largest-scale modes. This growth is broadband and increasingly concentrated toward the highest resolved wavenumbers.
Figure \ref{fig:dod_spectrum}(b) presents the temporal evolution of the total adjoint energy, defined as the spatial $L^{2}$ norm of the adjoint velocity field. The flow exhibits rapid growth of the ensemble-averaged adjoint energy during backward integration.

Near $\tau = 0$ or $t=T$, both quantities decay exponentially in time because diffusion dominates for a broadband perturbation.
After the initial transient region (for a time period around $0.3$), the red solid curve shows that the averaged adjoint energy, or variance $\overline{\|\boldsymbol{u}^{\dagger} \|^2}$, grows exponentially in backward time, confirming the chaotic divergence of trajectories. 
In contrast, the blue solid curve ($\|\overline{\boldsymbol{u}^{\dagger}}\|^2$) shows exponential decay even after the transient region.
This contrast highlights the fundamental difficulty in both optimization and the estimation of ensemble-averaged sensitivity: while the mean decays exponentially, the variance grows exponentially, implying that an exponentially increasing number of samples is required to maintain statistical convergence \citep{wang_wang_zaki_2022, wang2013forward}, as further indicated by 
the signal-to-noise ratio $\|\overline{\boldsymbol{u}^{\dagger}}\|^2/\overline{\| \boldsymbol{u}^{\dagger} \|^2}$ shown in black dashed curves.
Near the time $t=0$, or $\tau = T-t= 2$, the mean adjoint fields also exhibit growth in energy. This behavior arises from the randomized initial condition used in the forward simulation, which dominates the diffusive effects and leads to amplified adjoint sensitivity at later backward times.
The hatched region in figure~\ref{fig:dod_spectrum}(b) indicates the spin-up interval $\tau \geq 1.8$, corresponding to the first $T_{\mathrm{spin}}=0.2$ of forward time.

To illustrate the convergence of these statistics, we include lighter-colored lines in the background of panel $(b)$, representing the statistics calculated from different sizes of ensemble: ${50,100,200,400,1000,3000}$. As the sample size increases, these curves converge toward the solid lines. For chaotic systems, achieving full statistical convergence for the ensemble mean is difficult. For the current two-dimensional HIT system, the calculated $\|\overline{\boldsymbol{u}^{\dagger}}\|^2$ and $\overline{\| \boldsymbol{u}^{\dagger} \|^2}$ clearly converge to the thick lines computed using 3000 samples.

\begin{figure}[h]
    \centering
    \includegraphics[width=0.7\textwidth]{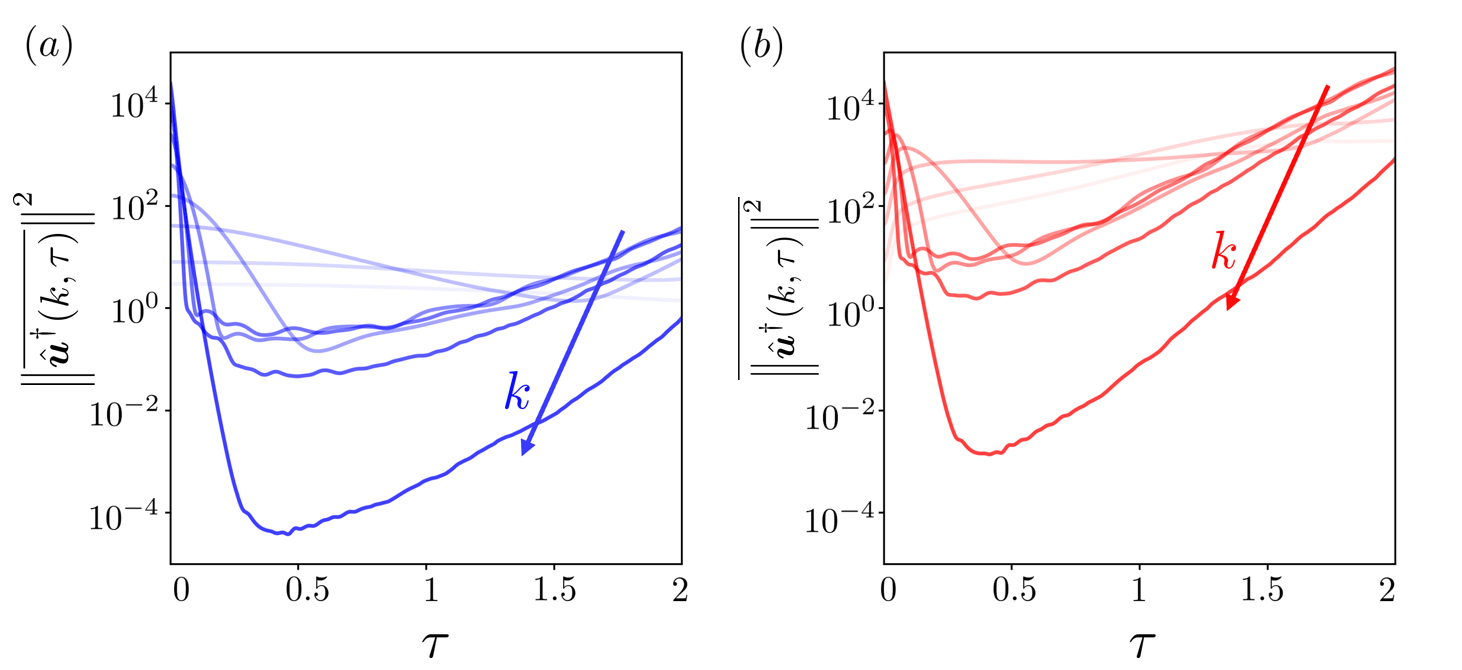}
    \caption{Shell-wise adjoint statistics for wavenumbers $k \in \{2,4,8,16,32,64,128,256\}$ in the decaying HIT case. Panel (a) shows $\lVert \overline{\hat{\boldsymbol{u}}^\dagger(k,\tau)} \rVert^{2}$, the energy of the ensemble-averaged adjoint in each wavenumber shell. Panel (b) shows $\overline{\lVert \hat{\boldsymbol{u}}^\dagger(k,\tau)\rVert^{2}}$, the mean per-trial shell energy (3000 samples). For both panels, lighter-colored curves correspond to smaller wavenumbers $k$.}

    \label{fig:adjoint_wavenumber}
\end{figure}

To pinpoint the non-uniform evolution of adjoint energy across wavenumber space, figure~\ref{fig:adjoint_wavenumber} shows the shell-wise energy of the ensemble-mean adjoint field, $E^{\dagger}(k,\tau)=\|\overline{\hat{\boldsymbol{u}}^{\dagger}}(k,\tau)\|^2$, and the corresponding ensemble-mean adjoint energy, $\overline{\|\hat{\boldsymbol{u}}^{\dagger}(k,\tau)\|^2}$, as functions of wavenumber. In decaying HIT, the large-scale adjoint components vary only weakly in backward time, whereas the high-wavenumber components amplify much more rapidly. Physically, this spectral analysis provides compelling evidence that the ``reverse butterfly effect" \citep{zaki2025arfm, wang_wang_zaki_2022} in the adjoint equation manifests primarily at small scales. The high-wavenumber components of the standard adjoint field are essentially noise that obscures the valuable low-wavenumber gradient information. This scale-dependent growth shows that small-scale adjoint sensitivities are dominated by incoherent backward amplification, making the corresponding gradient information unreliable for optimization. It also provides the physical motivation for the proposed spectral preconditioning: damping high-wavenumber gradient components suppresses this unstable adjoint growth while preserving the more reliable large-scale sensitivity.

The spectral analysis presented above reveals a fundamental barrier in adjoint-based data assimilation for turbulent flows. We observe a ``spectral catastrophe" where the adjoint energy cascades to small scales (high $k$) while the coherence of the sensitivity information simultaneously vanishes. 
This behavior is consistent with the broader physical picture of turbulence, in which coherent organization is primarily
associated with large-scale vortical structures, while smaller scales contain
stronger filamentary and less coherent fluctuations
\citep{jimenez2021collective}. Here, the adjoint counterpart of this
scale separation is the concentration of reproducible ensemble-mean
sensitivity at large scales and the rapid high-wavenumber growth observed
during backward integration.
This has critical implications for the optimization problem formulated in Section~\ref{sec:formulation}. In the standard adjoint approach, which relies on the $L^2$ inner product, the gradient is defined based on the total adjoint energy. Consequently, the optimization direction is dominated by the high-wavenumber components-precisely those scales that represent chaotic noise rather than physical sensitivity. This explains the poor reconstruction quality and convergence issues often encountered in standard adjoint data assimilation.
To address this issue, the framework provided here refines the gradient to capture the physically meaningful behavior. This requires a data-informed design of the preconditioner $\mathcal{G}$ that properly attenuates the weighting on high-wavenumber components while retaining the large-scale components for which the gradient remains reliable. This physical insight motivates the preconditioning strategies developed in the present work and points toward further approaches that explicitly exploit the spectral structure of adjoint energy growth.

\section{Conclusion}
\label{sec:conclusion}
In this study, we developed a mathematical framework for preconditioned adjoint data assimilation, specifically designed to address the challenges of reconstructing chaotic turbulent flows from sparse observations.
The method relies on a generalized definition of the inner product in the optimization, which introduces a spectral weighting kernel.
This formulation allows for the choice of control variables in a latent space, distinct from the physical velocity space, where the simulation and optimization are more effectively performed.
For the specific forms of the preconditioner being an algebraic or exponential weighting in wavenumber space, this approach can be interpreted as using the fractional derivative, or a sharpened version of the velocity as a control vector for the optimization.
Employing the preconditioner implicitly introduces a precursor simulation step, similar to solving a diffusion equation within the data assimilation loop, and helps regularize the ill-posed inverse problem.

\qw{In forced Kolmogorov flow, the best algebraic preconditioner reduces the normalized initial-condition velocity error from $0.545$ to $0.016$, providing the main quantitative result of this study.} \qw{Preconditioning also improves reconstruction in decaying HIT, where the exponential family performs best.} \qw{The difference in effectiveness is consistent with the separation between the energy-containing scales and the high-wavenumber components amplified by the adjoint.} \qw{Both flows support the same stabilization mechanism: selective suppression of incoherent high-wavenumber sensitivity, with the preferred kernel depending on the spectral characteristics of the flow.}
\qw{The improvement is robust to both observation resolution and assimilation-window length. Across all tested settings, the preconditioned methods outperform the standard adjoint, whose error remains above that of the trivial zero estimate even with the densest observations tested. The dependence on window length further indicates that this failure arises from spurious small-scale content in the gradient rather than insufficient information. In the decaying flow, late-time observations become insensitive to the small scales of the initial condition, thereby reducing their contribution to the reconstruction.}

In addition, we analyze the source of the effect for the preconditioned adjoint by statistically examining the adjoint sensitivity. We found that, under the standard formulation, adjoint energy exhibits exponential growth, particularly at small scales due to chaotic divergence. This growth is largely driven by incoherent small-scale fluctuations with a negligible signal-to-noise ratio. The preconditioned approach overcomes this limitation by explicitly suppressing these unreliable high-wavenumber components. By steering the gradient descent toward coherent large-scale structures where the sensitivity remains physically meaningful, the method avoids the ``spectral catastrophe'' inherent to the raw adjoint and enables stable data assimilation in chaotic regimes.

The current work is foundational and opens up directions for further studies. First, while this study demonstrates the efficacy of spectral preconditioning in two two-dimensional flows, the performance of the method in three-dimensional flows, where the forward energy cascade and vortex stretching dominate, remains to be investigated. Second, it remains an open question how the appropriate kernel family and its scale parameter depend on the characteristics of the underlying flow, which has the potential to further enable a dynamic, data-adaptive preconditioner that selects or updates both the kernel form and its scale during the assimilation process.

\section*{Acknowledgments}
This work was supported by the National Science Foundation under Grant No. 2332057, 2341393, and 2431610. Computational resources were provided through allocation MCH260115 from the Advanced Cyber infrastructure Coordination Ecosystem: Services \& Support (ACCESS) program.

\appendix
\section{Filtered adjoint equations}
\label{app:LES_adjoint}

We show here that defining the adjoint with respect to the modified inner product
\[
\langle \boldsymbol{a}, \boldsymbol{b} \rangle
=
\int_0^T \int_\Omega
\boldsymbol{a}^\top \mathcal{G}^{-1} \boldsymbol{b}
\, d\Omega \, dt
\]
naturally leads to a filtered form of the adjoint equations.

Let $\tilde{\boldsymbol{u}}^\dagger = \mathcal{G}\boldsymbol{u}^\dagger$ and $\tilde{p}^\dagger = \mathcal{G}p^\dagger$ denote the filtered adjoint velocity and pressure. For any linear operator $\mathcal{N}$, we obtain,
\begin{equation}
\begin{aligned}
\left\langle \mathcal{N}\boldsymbol{u}, \tilde{\boldsymbol{u}}^{\dagger} \right\rangle
&=
\int_0^T \int_{\Omega}
   (\mathcal{N}\boldsymbol{u})^{\top} \mathcal{G}^{-1} \tilde{\boldsymbol{u}}^{\dagger}
   \, d\Omega\, dt \\
&=
\int_0^T \int_{\Omega}
   (\mathcal{N}\boldsymbol{u})^{\top} \boldsymbol{u}^{\dagger}
   \, d\Omega\, dt \\
&=
\int_0^T \int_{\Omega}
   \boldsymbol{u}^{\top} \mathcal{N}^{\dagger}\boldsymbol{u}^{\dagger}
   \, d\Omega\, dt \\
&=
\int_0^T \int_{\Omega}
   \boldsymbol{u}^{\top} \mathcal{G}^{-1}
   \mathcal{G}\big(\mathcal{N}^{\dagger}\boldsymbol{u}^{\dagger}\big)
   \, d\Omega\, dt \\
&=
\left\langle
\boldsymbol{u},
\widetilde{\mathcal{N}^{\dagger}\boldsymbol{u}^{\dagger}}
\right\rangle.
\end{aligned}
\label{eq:duality}
\end{equation}

Thus, the adjoint operator with respect to the modified inner product corresponds to the filtered version of the classical adjoint operator.

Since the filter $\mathcal{G}$ is assumed to be symmetric and translation-equivariant in space, it commutes with spatial and temporal derivatives. Consequently, the filtered adjoint equations follow directly from the standard adjoint system.

Furthermore, the Fréchet derivative of the cost functional satisfies,
\begin{equation}
\label{eq:J_filtered}
\delta \mathcal{J}
=
\left[
\frac{\partial \mathcal{J}}{\partial \boldsymbol{u}},
\;\delta \boldsymbol{u}
\right]
=
\left\langle
\mathcal{G}\frac{\partial \mathcal{J}}{\partial \boldsymbol{u}},
\;\delta \boldsymbol{u}
\right\rangle
=
\left\langle
\widetilde{\frac{\partial \mathcal{J}}{\partial \boldsymbol{u}}},
\;\delta \boldsymbol{u}
\right\rangle.
\end{equation}
Since $\mathcal{G}$ represents filter, the derivative defined by the weighted inner product $\left\langle \cdot,\cdot \right\rangle$ is exactly the filtered version of that from the unweighted inner product $[\cdot,\cdot ]$.


Repeating the adjoint derivation using the modified inner product and the Lagrange multiplier $\tilde{\boldsymbol{u}}^\dagger$ and $\tilde{p}^{\dagger}$ yields the filtered adjoint system
\begin{equation}
\label{eq:filtered}
\begin{aligned}
\frac{\partial \tilde{\boldsymbol{u}}^\dagger}{\partial (-t)}
 + \widetilde{\nabla \boldsymbol{u}\cdot\boldsymbol{u}^\dagger}
 - \widetilde{(\boldsymbol{u}\cdot\nabla)\boldsymbol{u}^\dagger}
 &= \nabla \tilde{p}^\dagger
    + \nu \nabla^2 \tilde{\boldsymbol{u}}^\dagger
    +  \mathcal{G}\left[\mathcal{M}^{\top}\left(\mathcal{M}(\boldsymbol{u}) - \boldsymbol{m}\right)\right] - \delta(t)\;\lambda \nabla^2\tilde{\boldsymbol{u}}_0, \\
\nabla\cdot\tilde{\boldsymbol{u}}^\dagger
&=0.
\end{aligned}
\end{equation}

These equations are precisely the filtered form of the classical adjoint system \eqref{eq:AANS}. The nonlinear terms appear in filtered form, e.g.
$\widetilde{\nabla \boldsymbol{u}\cdot\boldsymbol{u}^\dagger}$ and
$\widetilde{(\boldsymbol{u}\cdot\nabla)\boldsymbol{u}^\dagger}$,
and would in general require a closure model if one were to express them solely in terms of the filtered adjoint variable $\tilde{\boldsymbol{u}}^\dagger$.

\section{Numerical Verification of the Control-Variable / Preconditioner Equivalence}
\label{app:cv_equivalence}

\begin{figure}[ht]
    \centering
    \includegraphics[width=0.8\textwidth]{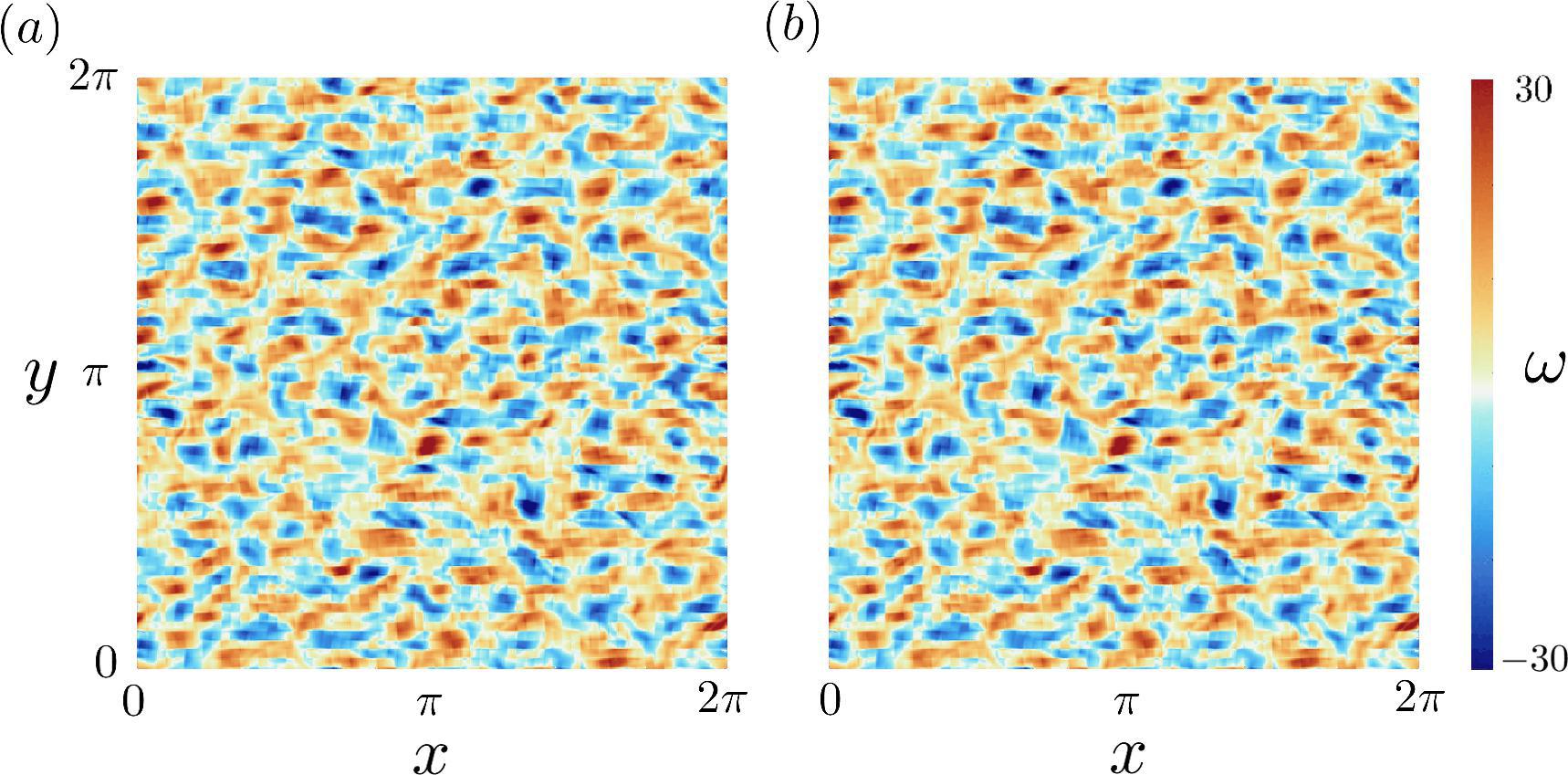}
    \caption{Comparison of initial condition reconstruction as a validation case. (a) Direct reconstruction using vorticity $\omega_0$ as control variable. (b) Reconstruction with preconditioner with Fourier multiplier $\hat{G} =1/k^2$.}
    \label{fig:omega_comparison}
\end{figure}

As a sanity check of the equivalence established in Section~\ref{sec:results_construction}, we compare the data assimilation using a preconditioner with Fourier multiplier $\hat{G}(k) = 1/k^2$ against the result obtained with the vorticity $\omega_0$ as the control variable. The left panel of figure~\ref{fig:omega_comparison} presents the reconstruction from vorticity-based control, while the right panel shows the result produced by applying spectral preconditioning to the velocity field using $k$-scaled modes. After $N=100$ optimization iterations, the two reconstructed initial conditions are visually indistinguishable, confirming the equivalence of the two approaches.

\section{Adjoint sensitivity analysis in 2D Kolmogorov Flow}
\label{app:dod_kolmogorov}
\begin{figure}[h]
\centering
\includegraphics[width=0.55\textwidth]{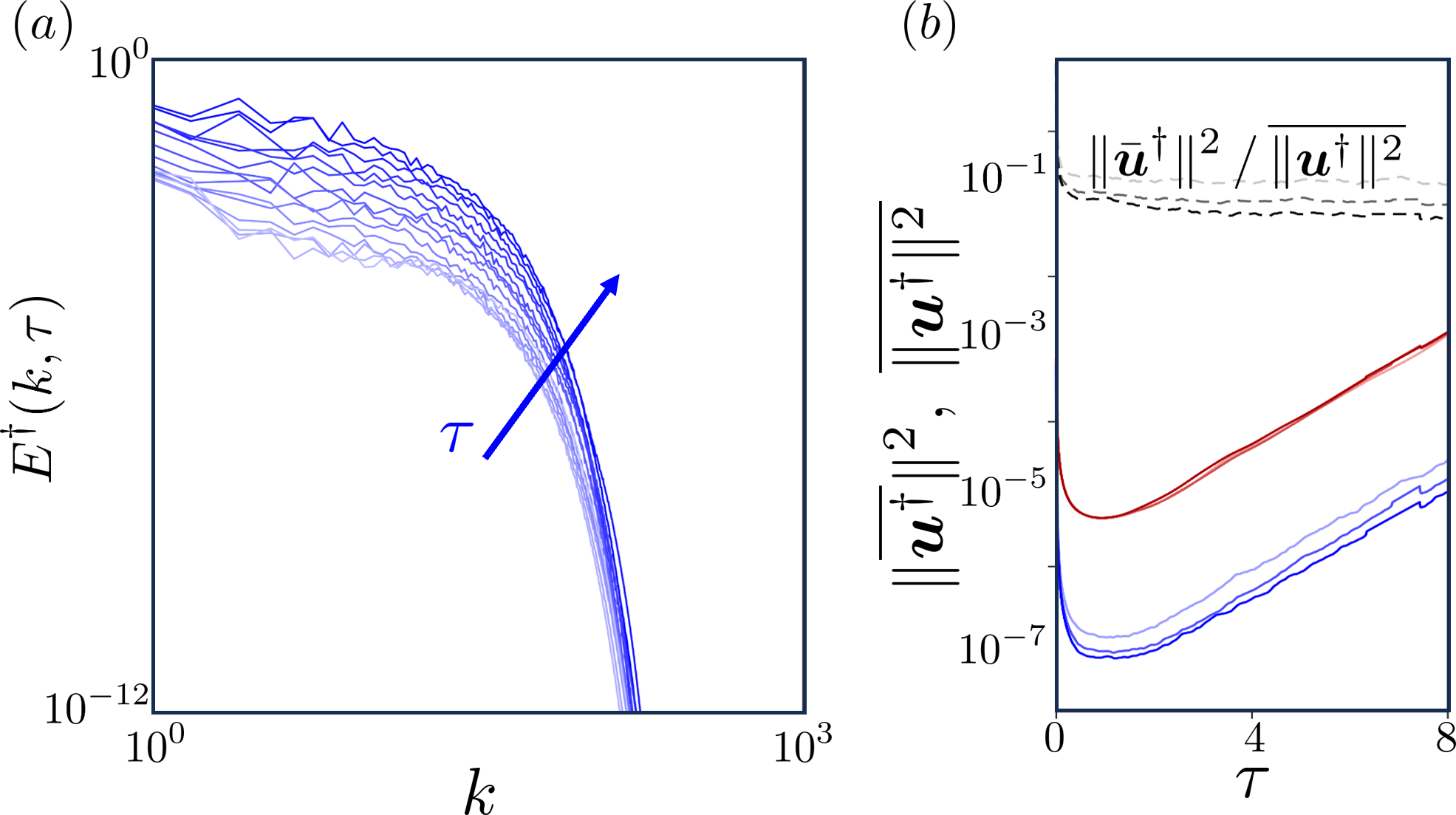}
\caption{Ensemble-averaged adjoint statistics from two-dimensional Kolmogorov flow, shown in backward time $\tau=T-t$. (a) The shell-averaged energy spectrum $E^{\dagger}(k,\tau)$, with darker shades indicating later backward times. (b) The integrated energies $\|\overline{\boldsymbol{u}^{\dagger}}\|^2$, $\overline{\|\boldsymbol{u}^{\dagger}\|^2}$, and their ratio; from lighter to darker curves, the ensemble subset sizes are $\{50,100,150\}$.}
\label{fig:dod_spectrum_kolm}
\end{figure}

We repeat the ensemble-based adjoint impulse-response analysis for the forced Kolmogorov flow, with the time window set to $T=8$. Figure~\ref{fig:dod_spectrum_kolm} reports the resulting statistics as functions of the backward time $\tau=T-t$.
Figure~\ref{fig:dod_spectrum_kolm}(a) shows that, as in decaying HIT, the Kolmogorov-flow adjoint energy evolves non-uniformly across wavenumbers during backward integration, with smaller-scale components amplifying more rapidly than the largest-scale modes. Unlike HIT, where this growth occurs over a wide range of wavenumbers, the Kolmogorov-flow adjoint spectrum shows a more localized amplification over a limited range of intermediate wavenumbers, while remaining comparatively steep at higher wavenumbers.
Figure~\ref{fig:dod_spectrum_kolm}(b) shows the corresponding temporal evolution of the total and ensemble-mean adjoint energies. As in HIT, the ensemble-averaged adjoint energy $\overline{\|\boldsymbol{u}^{\dagger}\|^2}$ grows exponentially in backward time, while the energy of the ensemble-mean adjoint field $\|\overline{\boldsymbol{u}^{\dagger}}\|^2$ and their ratio decrease; however, the overall growth is weaker than in decaying HIT.

The lighter background curves report ensemble subsets of size $\{50,100,150\}$. The total adjoint energy $\overline{\|\boldsymbol{u}^{\dagger}\|^2}$ is converged to within a few percent by $150$ samples. In contrast, the ensemble-mean energy $\|\overline{\boldsymbol{u}^{\dagger}}\|^2$ and the signal-to-noise ratio continue to decrease with sample size: for an incoherent adjoint field, the ensemble mean scales as $1/N$, so these quantities do not plateau, consistent with the difficulty of converging the ensemble mean in chaotic systems discussed in Section.~\ref{sec:stats_adjoint}. Thus, both flows exhibit the same qualitative pattern: the total adjoint energy grows in a scale-dependent manner during backward integration, while the coherent ensemble-mean component remains small. This shared behavior shows that the spectral motivation for the proposed preconditioner is not specific to decaying HIT.

\bibliography{references}
\end{document}